\documentclass{elsart}
\usepackage{amsmath}
\usepackage{amssymb}
\usepackage{amsbsy}
\usepackage{amsfonts}
\usepackage{graphics}
\usepackage{latexsym}
\usepackage{epsfig}
\usepackage{epsf}

\newcommand{\eq}{\begin{equation}}
\newcommand{\ee}{\end{equation}}
\newcommand{\ea}{\begin{eqnarray}}
\newcommand{\eea}{\end{eqnarray}}
\newcommand{\cW}{\mathcal{W}}
\newcommand{\cC}{\mathcal C}
\newcommand{\cO}{\mathcal O}

\begin{document}
\vspace{-1cm}
\begin{flushleft}
{DESY 03-162}\\
{ITEP-LAT/2002-28}\\
{KANAZAWA-03-07}\\
\end{flushleft}

\begin{frontmatter}

\title{Dynamics of Monopoles and Flux Tubes in 
%Two-Flavor Dynamical QCD\\[-1.5em]}
Two-Flavor Dynamical QCD}
\vspace{-0.3cm}

\author[KNZW,IHP,ITP]{V.G. Bornyakov},
\author[IPHU]{H. Ichie}\footnote{Present address:
Institute for Theoretical Physics, Kanazawa University, Kanazawa 920-1192, 
Japan},
\author[KNZW]{Y. Koma}\footnote{Present address:
Max-Plank-Institut f\"ur Physik, D-80805 M\"unchen, Germany},
\author[KNZW]{Y. Mori},
\author[KNZW]{Y. Nakamura},
\author[NI]{D. Pleiter},
\author[ITP]{M.I. Polikarpov},
\author[NI,DE]{G. Schierholz},
\author[NI,FUB]{T. Streuer},
\author[ZU]{H. St\"uben} and 
\author[KNZW]{T. Suzuki}

\vspace{0.15cm}

{-- DIK {\it Collaboration} --} 

\vspace{0.25cm}

\address[KNZW]{Institute for Theoretical Physics,
Kanazawa University, Kanazawa 920-1192, Japan}
\address[IHP]{Institute for High Energy Physics IHEP, RU-142284 Protvino,
  Russia} 
\address[ITP]{Institute of Theoretical and  Experimental
Physics ITEP, RU-117259 Moscow, Russia}
\address[IPHU]{Institut f\"ur  Physik, Humboldt-Universit\"at zu Berlin,
D-10115 Berlin, Germany}
\address[NI]{John von Neumann-Institut f\"ur Computing NIC,\\
Deutsches Elektronen-Synchrotron DESY,
D-15738 Zeuthen, Germany}
\address[DE]{Deutsches Elektronen-Synchrotron DESY,
D-22603 Hamburg, Germany}
\address[FUB]{Institut f\"ur Theoretische Physik, Freie Universit\"at Berlin,
D-14196 Berlin, Germany}
\address[ZU]{Konrad-Zuse-Zentrum f\"ur Informationstechnik Berlin ZIB,
D-14195 Berlin, Germany}

\date{ }
%---------------------------------------------------------------

\begin{abstract}
We investigate the confining properties of the QCD vacuum with $N_f=2$ 
flavors of dynamical quarks, and compare the results with the properties of 
the quenched theory. We use non-perturbatively $\mathcal{O}(a)$ improved Wilson
fermions to keep cut-off effects small. We focus on color magnetic monopoles.
Among the quantities we study are the monopole density and the monopole
screening length, the static potential and the profile of the color electric
flux tube. We furthermore derive the low-energy effective monopole action.
Marked differences between the quenched and dynamical vacuum are found.
\end{abstract}
\end{frontmatter}

\pagebreak

\section{Introduction}

The dynamics of the QCD vacuum, and color confinement in particular, becomes 
more transparent and amenable to quantitative investigation in the maximally 
abelian gauge (MAG)~\cite{tHooft,klsw}. In this gauge the relevant degrees of 
freedom are color electric charges, color magnetic monopoles, `photons' and 
`gluons'~\cite{ksw}. The latter appear to become 
massive~\cite{suganuma,vborn2} due to a yet 
unresolved mechanism, resulting in an abelian effective theory at large 
distances. There is evidence that the monopoles condense in the low temperature
phase of the theory~\cite{klsw,condense}, causing a dual Meissner effect which 
constricts the color electric fields into flux tubes, in accord 
with the dual superconductor picture of confinement. 

The dynamics of monopoles has been studied in detail in quenched lattice 
simulations. It turns out that in the MAG the string tension is accounted for 
almost entirely by the monopole part of the abelian projected gauge 
field~\cite{monst,bbms}, and that the low-energy effective monopole action is 
able to reproduce both the string tension and the low-lying glueball 
masses~\cite{ts}. Furthermore, many of the non-perturbative features of the 
vacuum, such as the topological charge density~\cite{hart,bs,bcp} and 
spontaneous chiral symmetry breaking~\cite{xsb}, can be traced back to 
monopole excitations.

Very little is known about the dynamics of monopoles in the full theory. So 
far the investigations have concentrated mainly on the static potential. While
the effect of sea quarks is clearly visible at short distances, even for 
relatively heavy quark masses~\cite{allt,SESA96,CPPA99}, no significant
changes have been observed in the long-range behavior of the potential and the
string tension. In contrast, the critical temperature of the chiral phase 
transition was found to depend noticeably on the mass of the dynamical 
quarks~\cite{kars1,ukaw1,dik}, which indicates that sea quarks have a visible 
effect on the non-perturbative properties of the vacuum as well.

It will be interesting now to see how the microscopic properties of the vacuum 
react to the introduction of dynamical color electric charges. In this paper we
shall study the effect of sea quarks on the dynamics of monopoles and the
confining potential, and on the effective monopole action. The paper is 
organized as follows. In Section 2 we present the details of our simulations,
as well as the gauge fixing procedure and abelian projection. In Section 3 we
discuss the gross properties of the vacuum, such as the monopole density and
the magnetic screening length, and the static potential. Furthermore, the 
problem of Gribov copies is addressed. Section 4 is devoted to a detailed
study of the static and dynamical properties of the color electric flux tube.
In Section 5 we derive the effective monopole action, employing an extended 
Swendsen method~\cite{Shiba:1995pu}. Finally, in Section 6 we conclude. 
Preliminary results of this work have been reported in Ref.~\cite{born2}.

\section{Simulation details}

Our studies are based on gauge field configurations with $N_f = 2$ flavors
of dynamical quarks generated by the QCDSF--UKQCD collaboration, using the 
Wilson gauge field action and non-perturbatively $O(a)$ improved Wilson 
fermions~\cite{booth1}: 
\eq
S_F = S_F^{(0)} - \frac{{\rm i}}{2}\kappa g\, c_{SW} a^5 \sum_s {\bar{\psi}(s)
\sigma_{\mu\nu} F_{\mu\nu}(s)\psi(s)},
\ee
where $S_F^{(0)}$ is the ordinary Wilson fermion action. Our data sample and 
run parameters are listed in Table 1. We will compare the results with the 
outcome of quenched simulations on lattices of similar size and lattice 
spacing. The parameters of our quenched runs are also given in Table 1.

We fix the MAG~\cite{brand1} by maximizing the functional
\eq \label{FU}
F[U]=\frac{1}{12\,V}
\sum_{s,\mu}{(|U_{11}(s,\mu)|^2+|U_{22}(s,\mu)|^2+|U_{33}(s,\mu)|^2)}
\ee
with respect to local gauge transformations $g$ of the lattice gauge
field, 
\eq
U(s,\mu) \to U^g(s,\mu) = g(s)^\dagger U(s,\mu)g(s+\hat \mu)\,.
\ee 
To do so, we use a simulated annealing (SA) algorithm~\cite{bali1}, in which 
the gauge transformed link variables $U^g$ are thermalized according to 
the probability distribution
\eq
p(U^g) = {\exp}\{F[U^g]/T\}\,,
\ee
where $T$ is an auxiliary `temperature' which is gradually decreased after 
every Monte Carlo sweep from $T=5$ to $T=0.04$. To do so, we use 7500 sweeps 
on the
$16^3\, 32$ lattice and 10000 sweeps on the $24^3\, 48$ lattice. Every sweep
consists of a heat bath update of each of the three $SU(2)$ subgroups of the 
link matrices. After the 
final temperature has been reached, several local gauge transformations are 
applied until $F[U]$ has attained its maximum value within machine precision. 
It is known~\cite{vborn,bbms} that the MAG is plagued by Gribov copies. This 
shows 
in the occurence of local maxima of $F[U]$. As a result, gauge non-invariant 
observables will, in general, depend on how close one gets to the global 
maximum. Although the SA algorithm performs much better than the iterative 
local maximization procedure used in $SU(3)$ gauge theory so far, it is not 
always able to find the global maximum. We shall estimate the systematic
error due to this problem in Section~3.

\begin{table}[t]
\begin{center}
\begin{tabular}{|c|c|l|l|l|c|l|} \hline
\multicolumn{7}{|c|}{$N_f=2$}\\
\hline \multicolumn{1}{|c|}{$\beta$} &\multicolumn{1}{|c|}{Volume} 
&\multicolumn{1}{|c|}{$\kappa$} &\multicolumn{1}{|c|}{$c_{SW}$}
&\multicolumn{1}{|c|}{$m_{\pi}/m_{\rho}$} 
&\multicolumn{1}{|c|}{$r_0/a$} & \multicolumn{1}{|c|}{$a$ [fm]} \\
\hline
5.20 & $16^3\, 32$ & 0.1355 & 2.0171 & 0.6014(73) & 5.04(4) & 0.0972(8) \\
5.25 & $24^3\, 48$ & 0.13575& 1.9603 & 0.6012(73) & 5.49(3)  & 0.0911(5) \\
5.29 & $24^3\, 48$ & 0.1355 & 1.9192 & 0.7029(49) & 5.57(2) & 0.0898(3) \\
5.29 & $16^3\, 32$ & 0.135  & 1.9192 & 0.7586(22) & 5.24(4)   & 0.0954(7) \\
5.29 & $16^3\, 32$ & 0.134  & 1.9192 & 0.8311(26) & 4.81(5) & 0.104(1) \\
\hline \hline \multicolumn{7}{|c|}{$N_f=0$}\\ \hline
\multicolumn{1}{|c|}{$\beta$} &\multicolumn{1}{|c|}{Volume} 
&\multicolumn{3}{|c|}{$ $} 
&\multicolumn{1}{|c|}{$r_0/a$} & \multicolumn{1}{|c|}{$a$ [fm]} \\
\hline
5.8  & $24^3\, 48$ &\multicolumn{3}{|c|}{$ $} & 3.67     & 0.137(2) \\
6.0  & $16^3\, 32$ &\multicolumn{3}{|c|}{$ $} & 5.37     & 0.091(1) \\
6.0  & $24^3\, 48$ &\multicolumn{3}{|c|}{$ $} & 5.37     & 0.091(1) \\
6.2  & $24^3\, 48$ &\multicolumn{3}{|c|}{$ $} & 7.38     & 0.068(2) \\
\hline
\end{tabular}
\vspace*{0.75cm}
\caption{Parameter values of our dynamical ($N_f=2$)~\cite{booth1} and 
quenched ($N_f=0$) gauge field configurations. The improvement coefficient 
$c_{SW}$ was computed in~\cite{janmer}. The quenched $r_0/a$ values have been 
taken from~\cite{Necco:2001xg}. We have used $r_0 = 0.5$ fm to set the scale.}
\end{center}
\end{table}

The functional $F[U]$ is invariant under local $U(1)\times U(1)$ gauge 
transformations and global Weyl transformations. From the (gauge fixed) 
$SU(3)$ link variables we derive abelian link variables~\cite{klsw,poli1}
\eq\label{ulink}
u(s,\mu) \equiv {\rm diag}\big(u_1(s,\mu), u_2(s,\mu), u_3(s,\mu)\big)\,, \quad
u_i(s,\mu) = \exp({\rm i}\, \theta_i(s,\mu)) 
\ee
with
\eq
\label{tlink}
\theta_i(s,\mu) = \arg(U_{ii}(s,\mu))-\frac{1}{3} \sum_{j=1}^3 
\arg(U_{jj}(s,\mu))\,\big|_{\,{\rm mod}\ 2\pi} \in [-\frac{4}{3}\pi,
\frac{4}{3}\pi]\,.
\ee
The abelian link variables $u(s,\mu)$ take values in $U(1)\times U(1)$. Under 
a general gauge transformation they transform as 
\eq
\begin{tabular}{c}
$u(s,\mu) \to d(s)^\dagger u(s,\mu) d(s+\hat{\mu})\,,$ \\
$d(s) ={\rm diag}\big(
\exp({\rm i}\,\alpha_1(s)),\exp({\rm i}\,\alpha_2(s)),
\exp(-{\rm i}(\alpha_1(s)+\alpha_2(s)))\big)\,.$
\end{tabular}
\ee

The monopole currents reside on links of the dual lattice and are defined by
\eq\label{current}
k_i(^*s,\mu) = \frac{1}{2\pi} \sum_{\raisebox{-0.25ex}{$\Box$}\, \in\, 
\partial f(s+\hat{\mu},\mu)}
{\rm arg} (u_i(\raisebox{-0.25ex}{$\Box$})) = 0,\pm 1,\pm 2 \, ,
\ee
where $u_i(\raisebox{-0.25ex}{$\Box$})$ is the product of abelian parallel 
transporters around the plaquette $\raisebox{-0.25ex}{$\Box$}$, and 
$f(s+\hat{\mu},\mu)$ is the elementary cube perpendicular to the 
$\mu$-direction with origin $s+\hat{\mu}$, with $\raisebox{-0.25ex}{$\Box$}$
inheriting its orientation from $\partial f(s+\hat{\mu},\mu)$. (Note that we 
differ here from the original and correct normalization of the monopole 
currents~\cite{ksw}.) The phases are
chosen such that
\eq
\sum_i {\rm arg} (u_i(\raisebox{-0.25ex}{$\Box$})) = 0\, \quad 
|{\rm arg} (u_i(\raisebox{-0.25ex}{$\Box$})) -{\rm arg} 
(u_j(\raisebox{-0.25ex}{$\Box$}))| \leq 2\pi\,.
\ee
Because of that,
\eq
\sum_{i=1}^3 k_i(^*s,\mu) = 0 \,. 
\ee

\section{Gross properties of the vacuum and static potential}

Let us first look at the global changes of the QCD vacuum upon introducing
dynamical color charges. Our smallest quark masses are of the order of the 
strange quark mass.    

\subsection*{Monopole density}

\begin{figure}[t]
\begin{center}
\epsfig{file=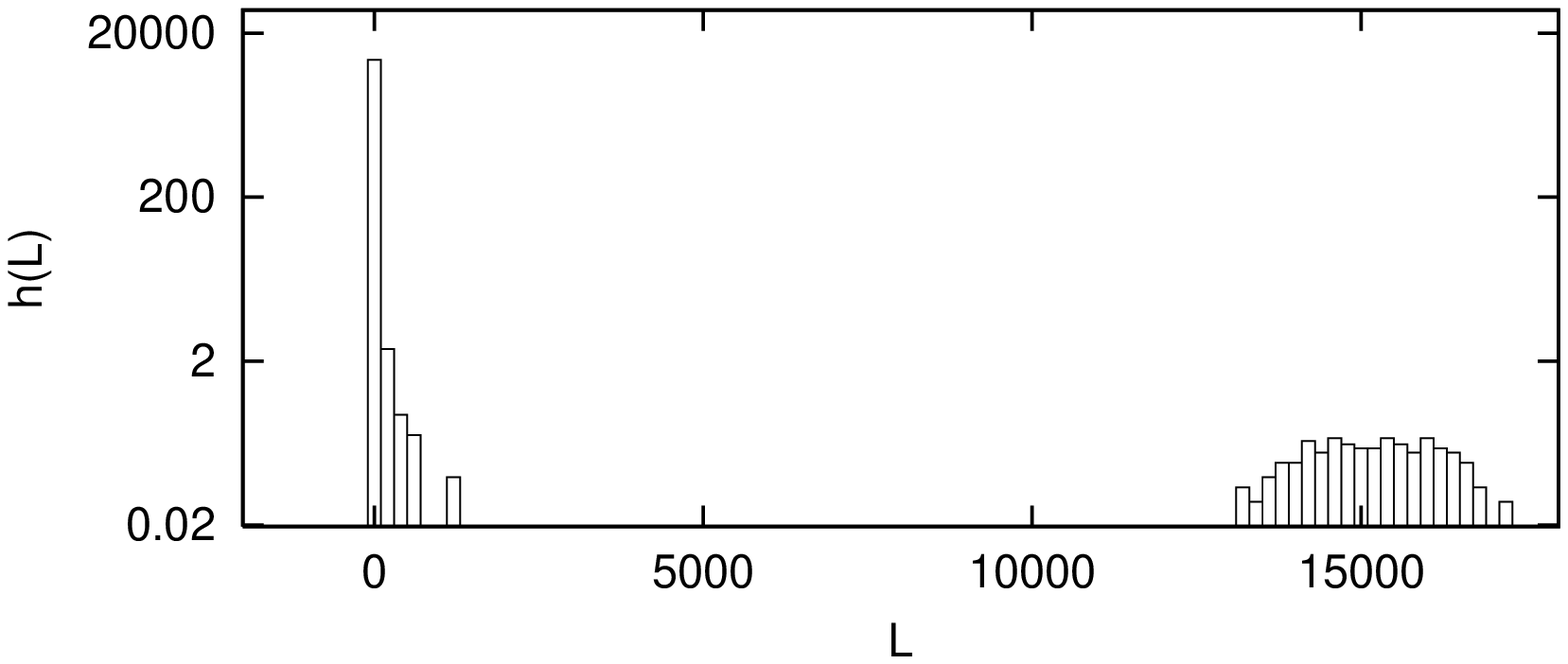,height=5.5cm,clip=}
\vspace*{0.25cm}
\caption{The histogram of closed monopole loops of length $L$ in full QCD on 
the $24^3\, 48$ lattice at $\beta=5.29$, $\kappa = 0.1355$. The bin size is 
200.} 
\label{abc}
\vspace*{1cm}
\hspace*{-0.4cm}\epsfig{file=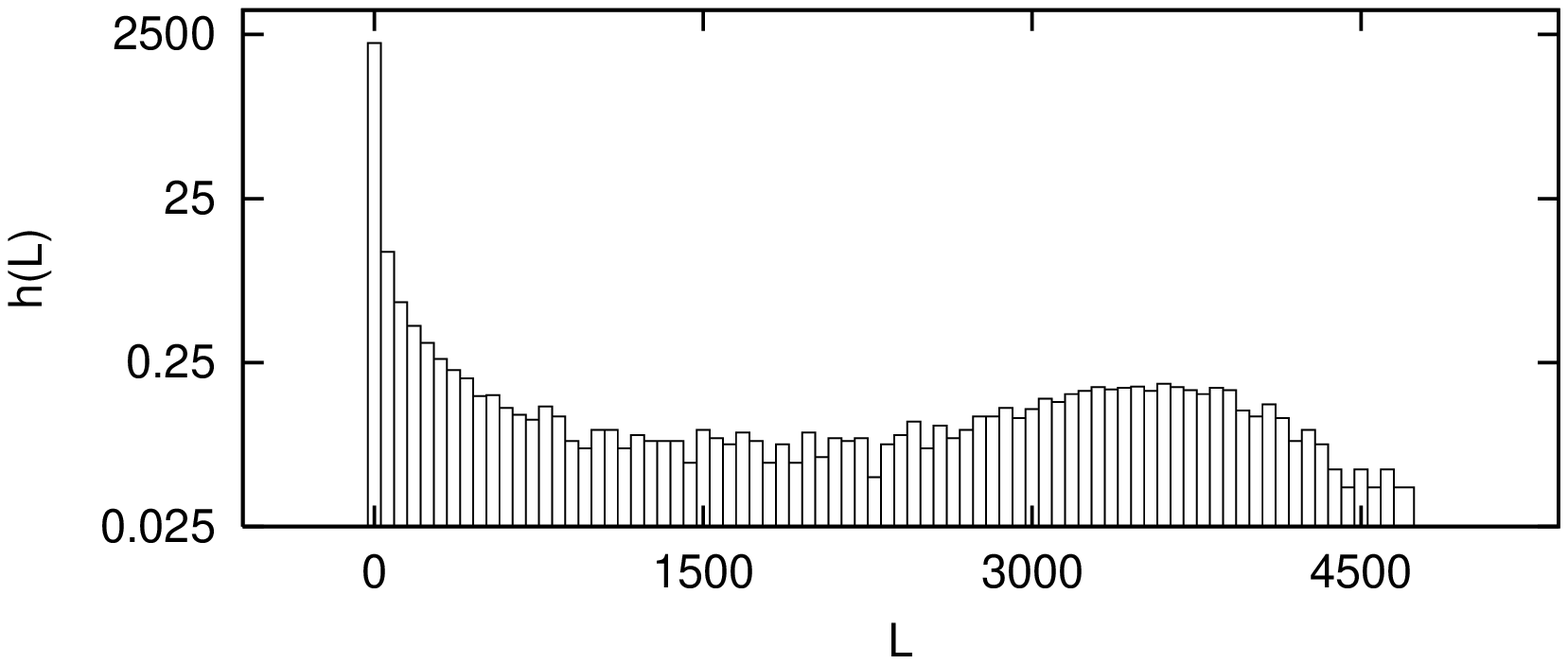,height=5.5cm,clip=}
\vspace*{0.25cm}
\caption{The same as in Fig.~\ref{abc}, but on the $16^3\, 32$ lattice at 
$\beta=5.29$, $\kappa = 0.135$. The bin size is 60.}
\label{spec16}
\vspace*{0.5cm}
\end{center}
\end{figure}

The monopole currents, which are conserved, form clusters of closed loops
on the dual lattice. In case of the pure $SU(2)$ gauge theory 
it was observed that these clusters fall into two different 
classes~\cite{kitahara,hart1}: `small' (ultraviolet) clusters which are
of limited extent in lattice units, and
`large' (infrared) clusters which percolate through the lattice and typically 
wrap around the boundaries. If the 
size of the lattice is large enough, a gap opens between the small and the 
large clusters, clearly separating the two. In general, each configuration 
accommodates at least one large cluster~\cite{hart1}. In Fig.~\ref{abc} we 
show the histogram $h(L)$ of monopole currents of length $L$ on the 
$24^3\, 48$ lattice. We observe two distinct clusters. For comparison we show 
the same quantity on the $16^3\, 32$ lattice in Fig.~\ref{spec16}. Both 
lattices have a similar lattice spacing and quark mass. On the smaller volume 
no gap is observed. For the long-distance properties of the vacuum all what 
matters is the existence of long, percolating monopole loops. Whether they
combine to one cluster or not is of secondary importance.

On smaller lattices we call a monopole cluster infrared if it forms the largest
cluster or if the monopole loop wraps around the boundary. It has been shown 
in the pure $SU(2)$ gauge theory~\cite{born1} that the corresponding monopole 
density does not depend on the lattice volume and scales properly, while the 
total monopole density diverges in the continuum limit.

\begin{figure}[t]
\begin{center}
\epsfxsize=10.0cm \epsfbox{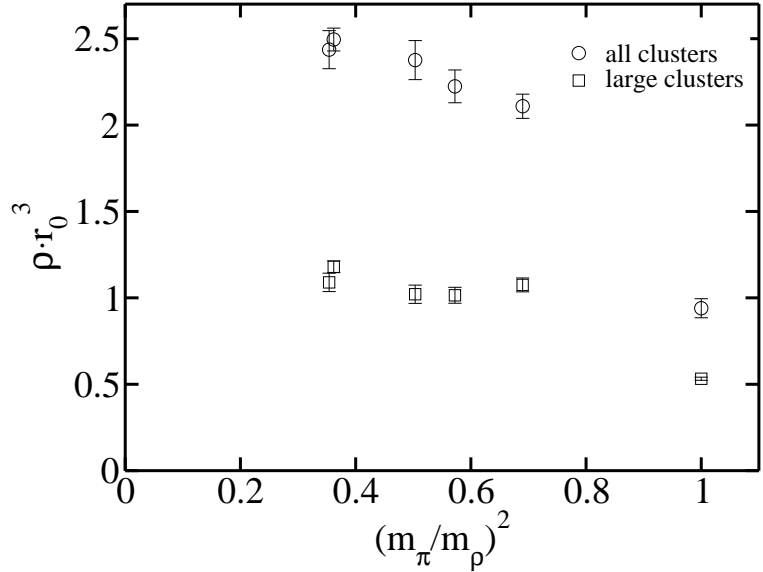} 
\caption{The monopole density in
full and quenched QCD. The quenched results refer to $\beta = 6.0$ and
are listed at $m_\pi/m_\rho = 1$.} \label{fig:rho}
\end{center}
\vspace*{0.25cm}
\end{figure}

We define the monopole density by
\eq
\rho=\frac{1}{12\,V}\sum_i\sum_{s,\mu}|k_i(^*s,\mu)|\,.
\ee
In Fig.~\ref{fig:rho} we show the total monopole density as well as the 
density of monopoles belonging to the infrared cluster. In the former case the
sum over $s, \mu$ extends over the full lattice, while in the latter case it 
extends over the links of the infrared clusters only. The quenched result is 
entered at $m_\pi/m_\rho = 1$. In the dynamical vacuum both densities are 
about a factor of two larger than in the quenched case. The total monopole
density appears to increase with decreasing quark mass, while the density of
infrared monopoles shows little variation, apart from the initial jump.

How can one explain the increase of the monopole density in the dynamical 
vacuum? It has been known for some time that monopoles are induced by 
(anti-)instantons~\cite{hart,bs}, at least partially. The fermion determinant 
introduces an attraction between instantons and anti-instantons, and the
force increases with decreasing quark mass~\cite{ahasen}. The effect is, very 
likely, that the vacuum becomes solidly packed with instantons and 
anti-instantons, while isolated instantons are suppressed. As a result the 
density of (anti-)instantons increases, and consequently the density of 
monopoles.

One should be aware that the total monopole density is not universal
but depends, in general, on the action chosen, and that after
integrating out the fermions the effective gluonic action may be more noisy 
than the Wilson gauge field action at the same lattice spacing. 
This would mainly affect the total monopole density, and to a lesser extent 
the infrared cluster. 

\subsection*{Static potential}

From the abelian projected link variables $u_i(s,\mu)$ we extract
the abelian static quark-antiquark-potential. In order to improve the overlap
with the ground state, we smear the space-like links according to
\eq
u_i(s,j) \rightarrow \alpha\, u_i(s,j) + \sum_{k \neq j} u_i(s,k)
u_i(s+\hat{k},j) u_i(s+\hat{\jmath},k)^\dagger \,.
\label{smear}
\ee
We apply 30 smearing sweeps with $\alpha=2$. The abelian
potential $V_{\rm ab}(R)$ is given by
\eq
\label{plateau}
V_{\rm ab}(R)=\lim_{T \rightarrow \infty}\, \log\Big(
{\frac{< W_{\rm ab}(R,T) >}{< W_{\rm ab}(R,T+1)>}}\Big) \,,
\ee
where
\eq\label{Wab}
W_{\rm ab}(R,T) = \frac{1}{3}\,\mbox{Re}\,\mbox{Tr}\, \cW_\cC \,, 
\quad \cW_\cC  = 
\prod_{s,\mu\, \in\, \cC} u(s,\mu)\,,
\ee
and $\cC$ is a (orientated) loop of spatial extent $R$ and temporal extent $T$.

\begin{figure}[t]
\begin{center}
\epsfig{file=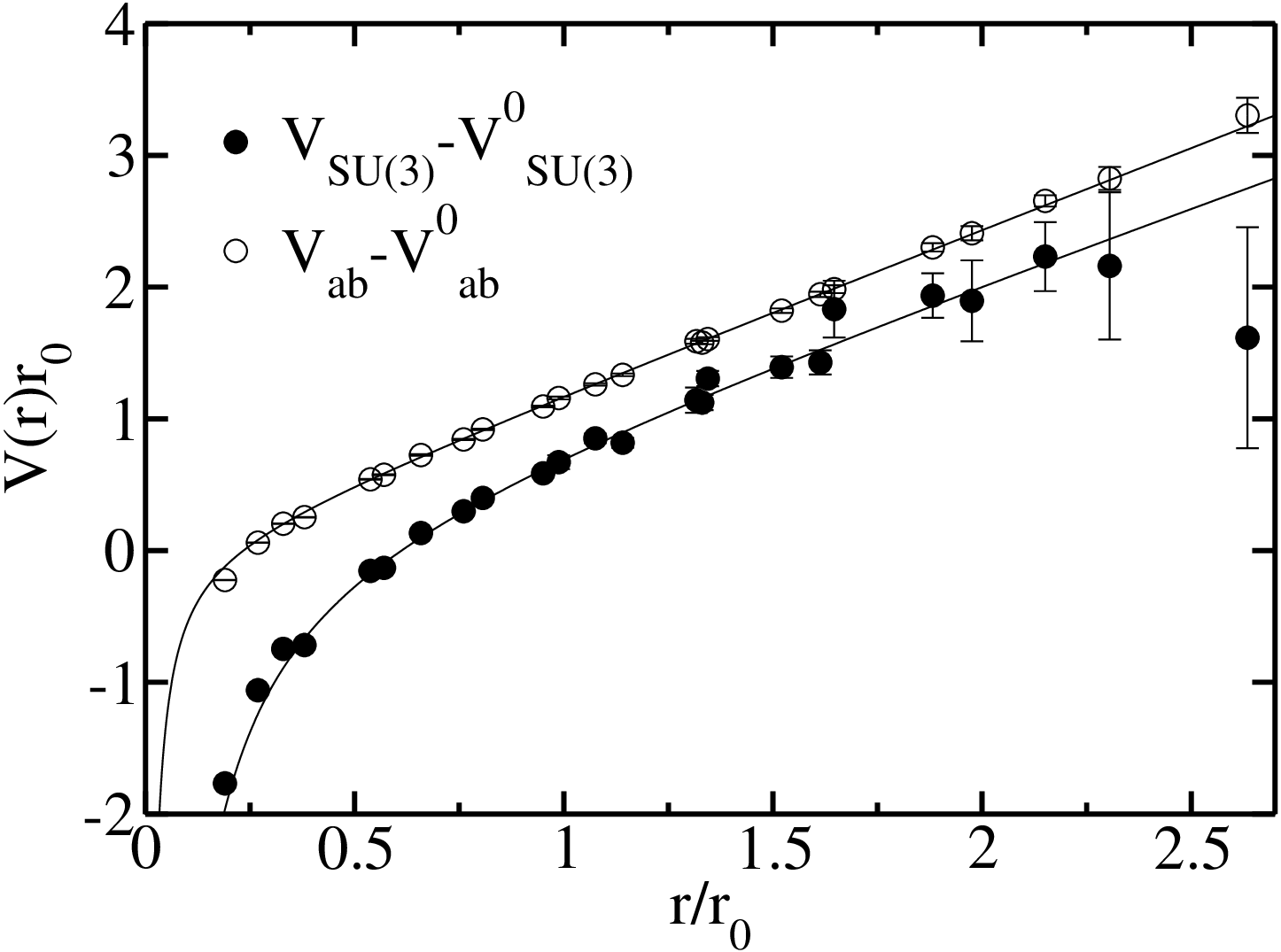,width=10cm,clip=}
\vspace*{0.25cm}
\caption{Comparison of the abelian and non-abelian potential in full QCD on
the $16^3\, 32$ lattice at $\beta=5.29$, $\kappa=0.135$. The solid lines are 
fits of the form (\ref{ansatz}).}
\label{pot_d1}
\vspace*{1cm}
\epsfig{file=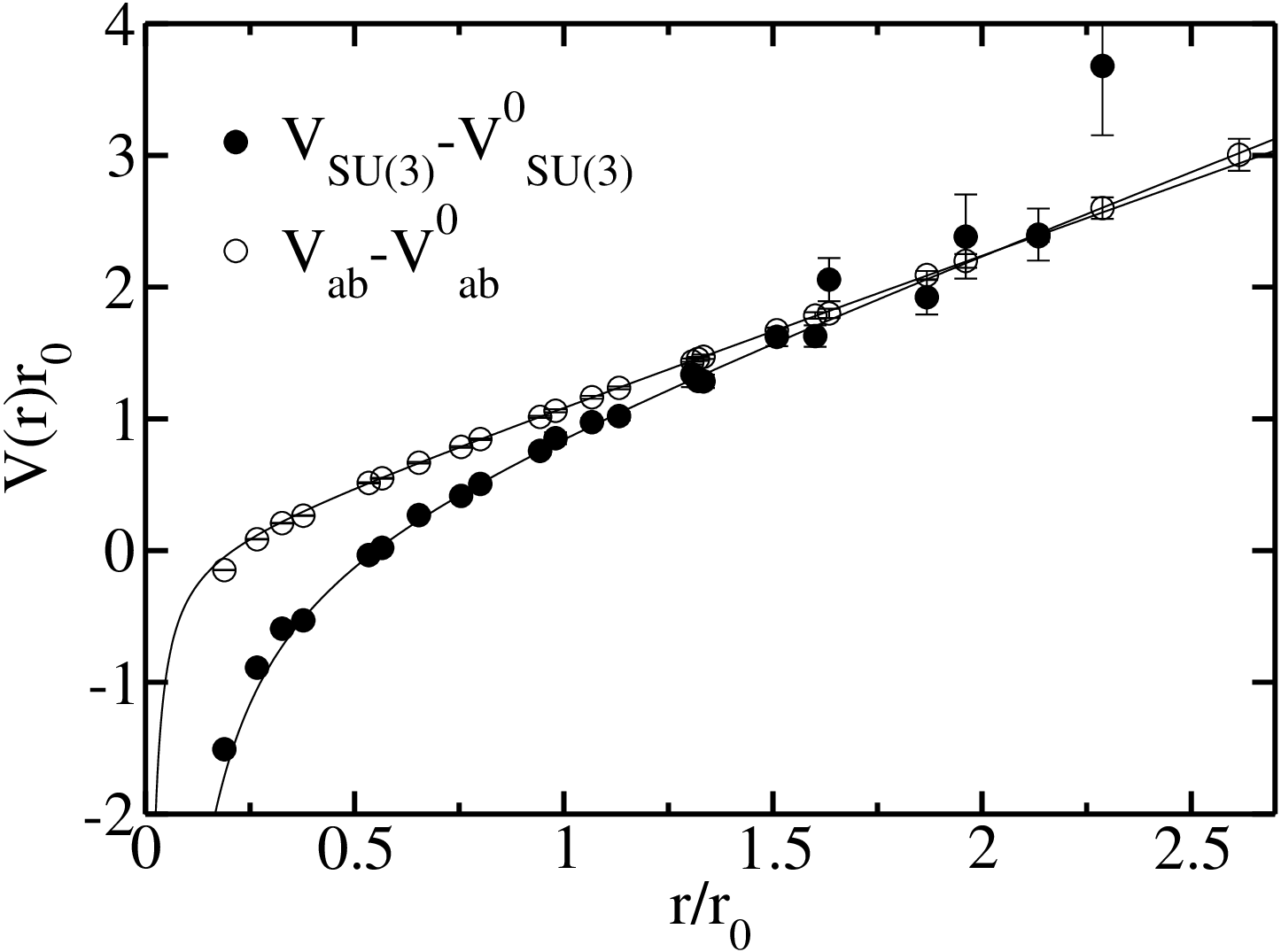,width=10cm,clip=} 
\vspace*{0.25cm}
\caption{Same as in Fig.~\ref{pot_q1}, but for the quenched theory at 
$\beta=6.0$.}
\label{pot_q1}
\vspace*{0.5cm}
\end{center}
\end{figure}

The ratio on the r.h.s. of (\ref{plateau}) reaches a plateau at $T=5$, so that
we will take $T=5$ throughout this section. We fit $V_{\rm ab}(r)$ by the 
ansatz
\eq
\label{ansatz}
V_{\rm ab}(r)=V^0_{\rm ab}+\sigma_{\rm ab} r-\frac{\alpha_{\rm ab}}{r} \,.
\ee
The potential was calculated for on-axis and off-axis directions 
$\hat{r}=1/\sqrt{2}\,(1,1,0)$ and $1/\sqrt{3}\,(1,1,1)$. At small $r$ 
rotational symmetry 
is broken on the lattice, and we exclude the first four data points from our 
fits. The non-abelian static potential was extracted from the non-abelian 
Wilson loop using a corresponding procedure. 

\begin{table}[t]
\begin{center}
\begin{tabular}{|c|c|c|c|c|c|}
\hline
$m_\pi/m_\rho$ & $\alpha_{\rm ab}$ & $\sigma_{\rm ab}/\sigma$ & 
$\sigma_{\rm mon}/\sigma_{\rm ab}$
& $\xi/r_0$ & $\sigma_{\rm ab}/\rho\, \xi$ \\
\hline
0.6014(73) & 0.12(1) &0.90(4) & 0.80(4) & 0.484(19) & 2.1(2) \\
0.7029(49) & 0.10(1) &0.96(3) & 0.87(3) & 0.466(26) & 2.6(3) \\
0.7586(22) & 0.11(1) &0.99(6) & 0.83(8) & 0.521(17) & 2.3(2) \\
0.8311(26) & 0.11(1) &0.99(6) & 0.88(5) & 0.482(17) & 2.5(2) \\
1     & 0.09(1) &0.83(3) & 0.84(3) & 0.662(34) & 3.2(3) \\
\hline
\end{tabular}
\vspace*{0.75cm}
\caption{The Coulomb term, the abelian and monopole part of the string 
tension, as well as
the monopole screening length in full and quenched QCD. The quenched result 
is shown in the last row and refers to the $16^3\, 32$ lattice at $\beta=6.0$.}
\end{center}
\vspace*{0.5cm}
\end{table}

In Figs.~\ref{pot_d1} and \ref{pot_q1} we show the abelian and non-abelian
static potential for some data set. The self-energy contributions have been 
subtracted. The ratios of abelian and non-abelian string tensions are given in
Table~2. As our determination of the non-abelian string tension was not 
accurate enough, we took these numbers from the literature: 
$\sqrt{\sigma} r_0  = 1.142(5)$ in full QCD~\cite{UKQCD} 
($\sqrt{\sigma} r_0$ depends only weakly on the
dynamical quark mass) and $\sqrt{\sigma} r_0 = 1.16(1)$ in the quenched
theory~\cite{QCDSF}.
The abelian string tension turns out to be very close to the non-abelian one in
full QCD, while in the quenched case it is noticeably smaller. But the ratio
of $\sigma_{\rm ab}$ to $\sigma$ may increase in the continuum 
limit~\cite{born1}.

\begin{figure}[t]
\begin{center}
\epsfig{file=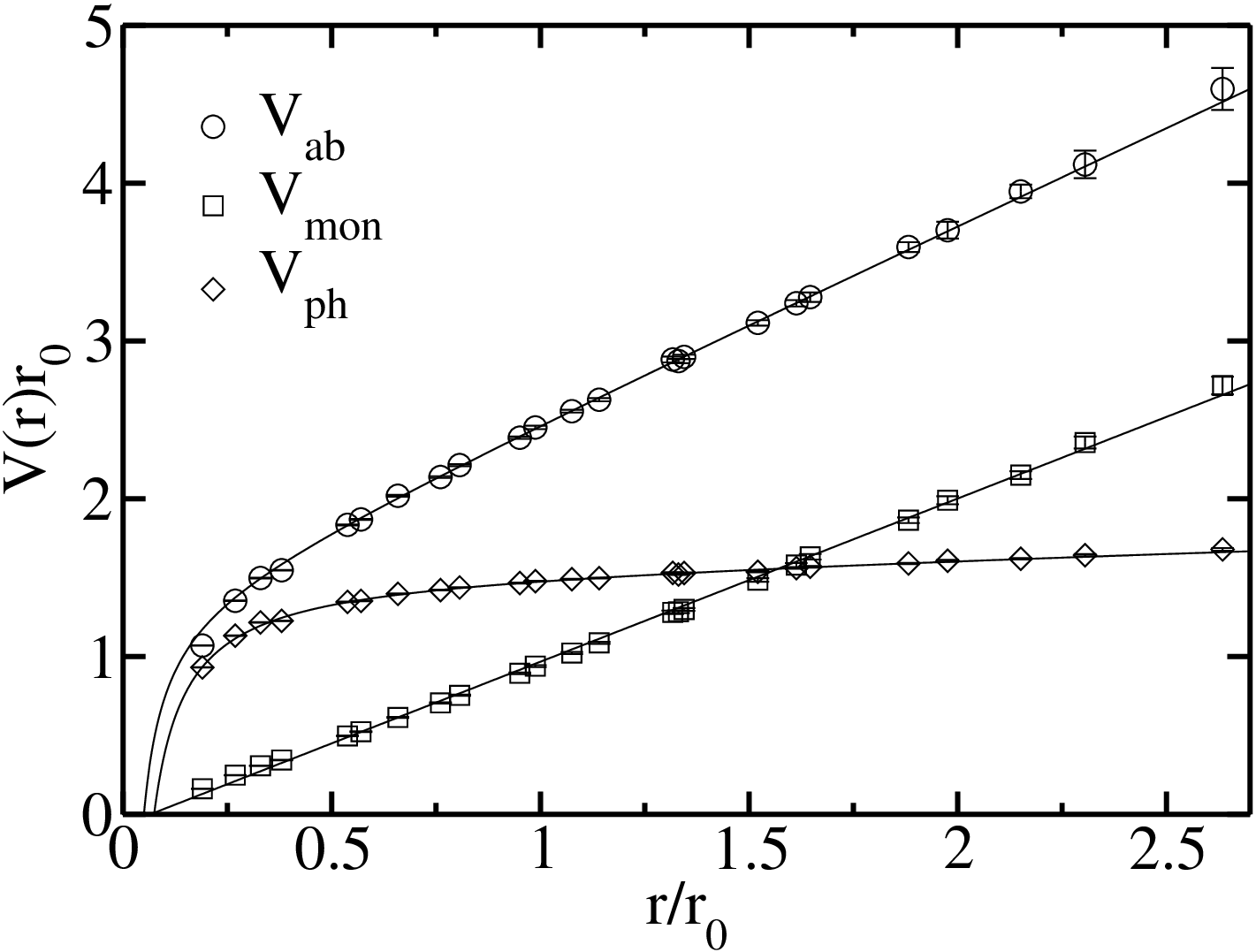,width=10cm,clip=} 
\vspace*{0.25cm}
\caption{Decomposition of the abelian potential into monopole and photon parts
on the $16^3\, 32$ lattice at $\beta=5.29$, $\kappa=0.135$ in full QCD. The 
solid lines are fits of the form (\ref{ansatz}).} 
\label{pot_d2}
\vspace*{1cm}
\epsfig{file=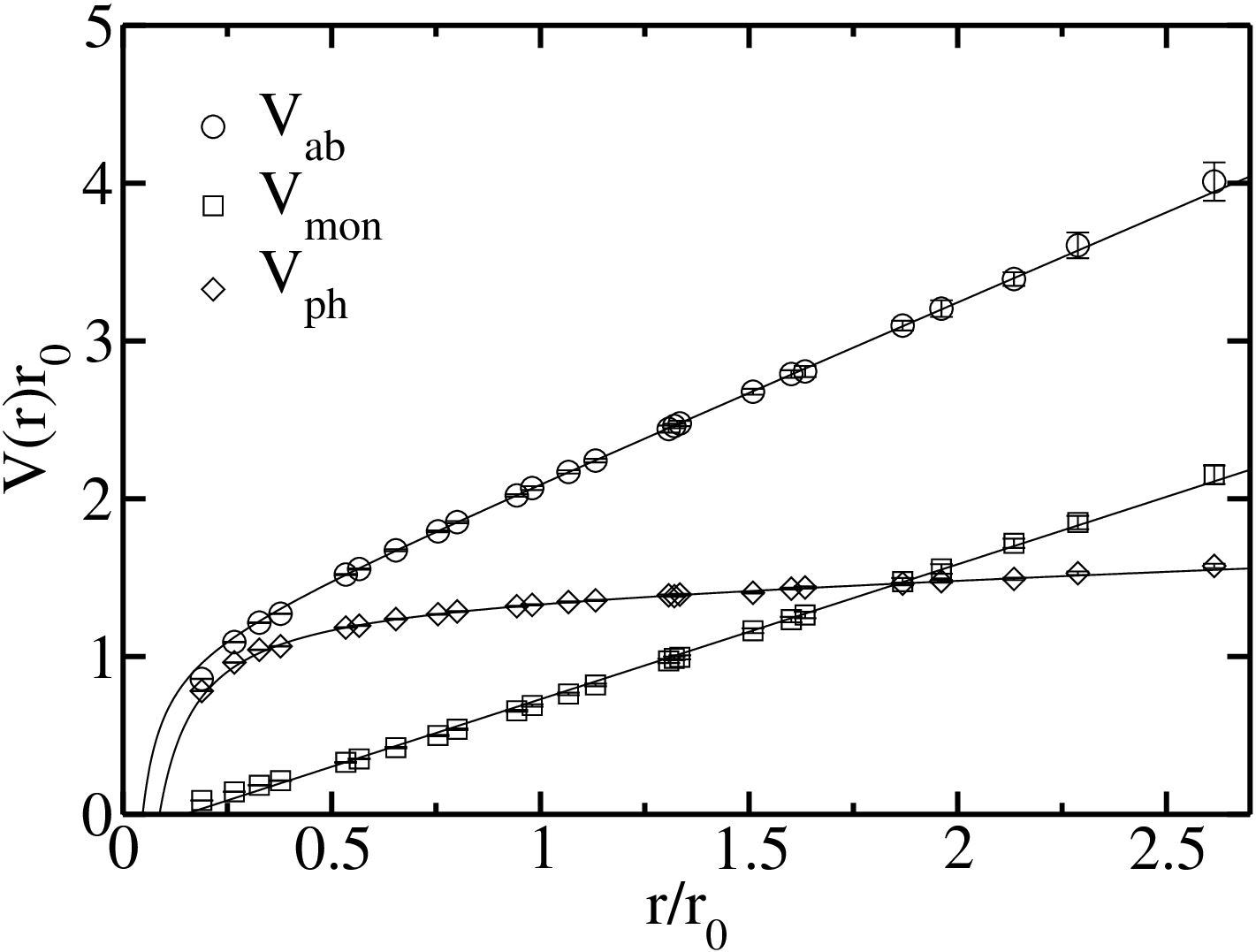,width=10cm,clip=}
\vspace*{0.25cm}
\caption{The same as in Fig.~\ref{pot_d2}, but for the quenched theory at
$\beta=6.0$.} 
\label{pot_q2}
\vspace*{0.5cm}
\end{center}
\end{figure}

The abelian link variables can be decomposed into a `singular' monopole part 
and a photon part according to the definition~\cite{smit1,sibasuzuki}:
\begin{gather}
\label{monph}
\theta_i(s,\mu) = \theta_i^{\rm mon}(s,\mu)+\theta_i^{\rm ph}(s,\mu)\,, \\
\theta_i^{\rm mon}(s,\mu)= 2 \pi \sum_{s'} D(s-s')\nabla^{(-)}_{\alpha}
m_i(s',\alpha,\mu)\,,
\label{montheta}
\end{gather}
where $D(s) = \Delta^{-1} (s)$ is the lattice Coulomb propagator,
$\nabla^{(-)}_{\mu}$ is the lattice backward derivative, and 
$m_i(s,\mu,\nu)$ counts the number of Dirac strings piercing the 
plaquette 
\eq
u_i(s,\mu,\nu) = u_i(s,\mu) u_i(s+\hat{\mu},\nu) 
u^\dagger_i(s+\hat{\nu},\mu) u^\dagger_i(s,\nu) \,.
\ee
If one computes $k_i(^*s,\mu)$ from 
$\theta_i^{\rm mon}(s,\mu)$ one recovers almost all monopole currents found
by the definition (\ref{current}), hence the notation monopole 
part~\cite{xsb}.

Similarly, from Wilson loops composed of the monopole (photon) 
part of the link variables one can derive the monopole (photon) contribution
to the static abelian potential. In Figs.~\ref{pot_d2} and \ref{pot_q2} we 
show both contributions. The string tension of the monopole part 
$\sigma_{\rm mon}$ is given in Table~2, while the photon contribution to the
potential is short-range. Within the uncertainties (of $O(a^2)$ corrections)
all string tensions are very similar, in the dynamical as well as in the 
quenched theory.

\begin{figure}
\begin{center}
\epsfxsize=10.cm \epsfbox{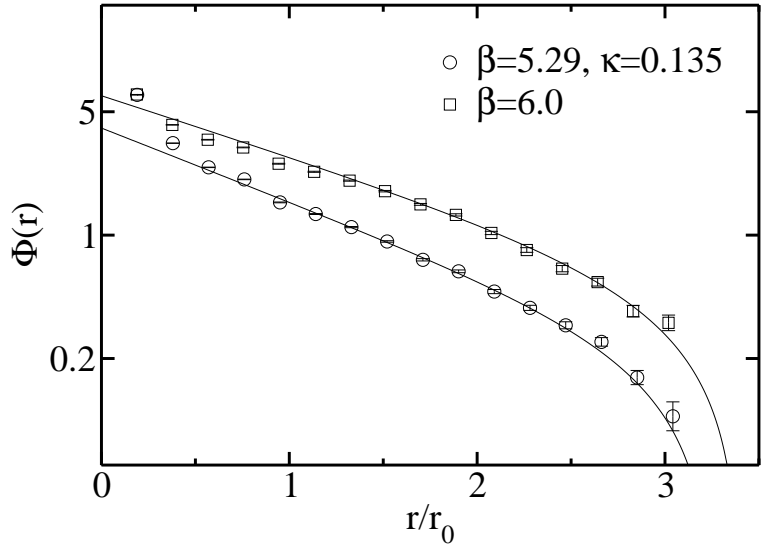} \caption{Data and fit of the 
magnetic flux on the $16^3\, 32$ lattice in full and quenched QCD.}
\label{screening}
\end{center}
\vspace*{0.5cm}
\end{figure}

The next quantity we looked at is the magnetic screening length $\xi$. This is 
defined by the exponential decay of the magnetic flux $\Phi(r)$ through a 
sphere of radius $r$ around the monopole. On a periodic lattice 
this can be written
\eq
\label{eqscreening}
\Phi(r) = \Phi_0 \exp\Big(-\frac{L}{2\xi}\Big) \sinh\Big(\frac{L-2r}
{2\xi}\Big)\,,
\ee  
where $L$ is the effective length of the box, which is taken to be a free 
parameter. 
Some numerical data are shown in Fig.~\ref{screening}, together with a fit of
eq.~(\ref{eqscreening}). The fitted values of $\xi$ for all data sets are 
given in Table~2. The length $L$ turns out to be slightly larger than the 
extent of the lattice, as expected. We notice that the screening length is 
about 30\% lower in
the dynamical vacuum as compared to the quenched case. This does not come 
unexpected. In a three-dimensional model of a monopole gas with 
screening~\cite{hart1} the abelian string tension turns out to be proportional
to the product of monopole density and screening length, i.e. $\sigma_{\rm ab}
\propto \rho\, \xi$. Though this model is an oversimplification of the 
underlying dynamics, it is in qualitative agreement with our findings.

\subsection*{Gribov copies}

\begin{figure}[t]
\vspace{.7cm}
\begin{center}
\epsfig{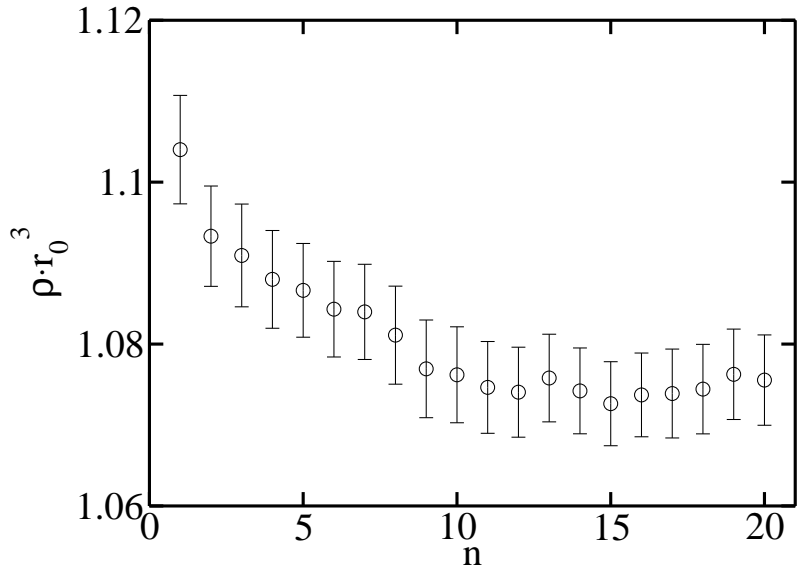} 
\vspace*{0.25cm}
\caption{The monopole density 
on the $16^3\, 32$ lattice at $\beta=5.29$, $\kappa=0.135$, and its dependence
on the number of Gribov copies.}
\label{rho_gc}
\vspace*{0.5cm}
\end{center}
\end{figure}

\begin{figure}[htbp]
\begin{center}
\epsfig{file=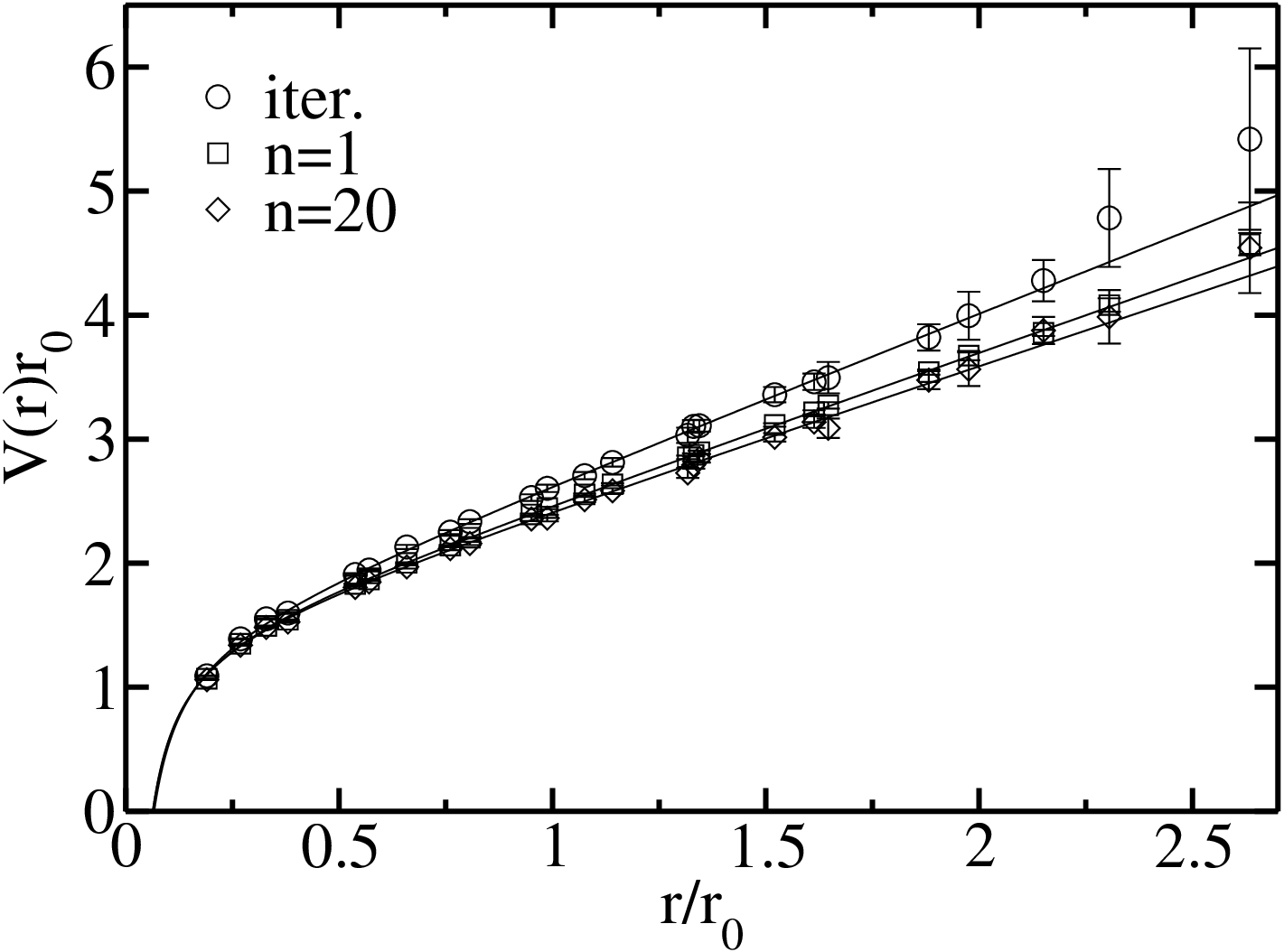,width=10cm,clip=} 
\vspace*{0.25cm}
\caption{The abelian static potential on the $16^3\, 32$ lattice at 
$\beta=5.29$, $\kappa=0.135$, and its dependence on the number of Gribov 
copies. Also shown is the result of the iterative gauge fixing.}
\label{pot_gc}
\vspace*{1.5cm}
\hspace*{-0.75cm} \epsfig{file=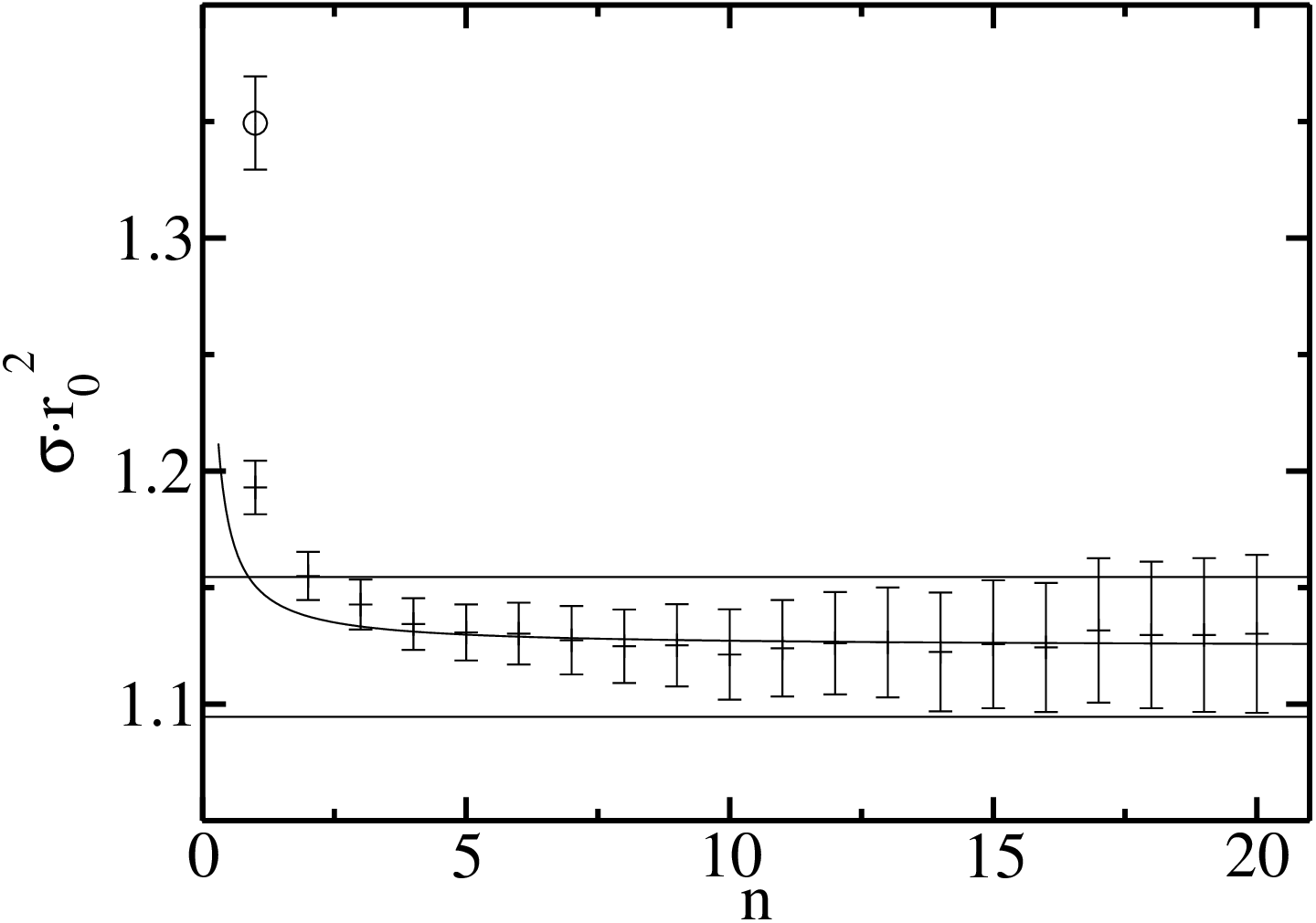,width=10.7cm,clip=} 
\vspace*{0.25cm}
\caption{The abelian string tension on the $16^3\, 32$ lattice at 
$\beta=5.29$, $\kappa=0.135$, and its dependence on the number of Gribov 
copies, together with the result of the iterative gauge fixing 
($\circ$).}
\label{st_gc}
\vspace*{0.5cm}
\end{center}
\end{figure}

We shall now try to quantify the error that is made by fixing to a local 
maximum of $F[U]$ instead of the global one. We follow the procedure 
suggested in~\cite{bbms}. The test runs are done on the $16^3\, 32$ lattice
at $\beta=5.29$, $\kappa=0.135$ using a total of $O(50)$ configurations.
We create 20 random gauge copies for each configuration, employing the SA
algorithm. Then we randomly pick $n$ gauge copies out of each set (of 20) and
select the copy with the highest value of $F[U]$ to compute our observable
${\mathcal O}$. Obviously the result will depend on $n$, and the true result
is obtained at $n \rightarrow \infty$. The approach to $n=\infty$ may be 
fitted by~\cite{Bornyakov:2000ig}
\eq
\langle {\mathcal O}\rangle(n) = \langle {\mathcal O}\rangle(\infty)
+\frac{\rm const.}{n}\,.
\ee
In Fig.~\ref{rho_gc} we show the monopole density as a function of $n$. We see
that $\rho$ reaches a plateau at $n \approx 10$. By taking only one gauge copy
into account one introduces a systematic error of the order of 3\%.
The effect of Gribov copies is slightly stronger in case of the abelian static 
potential, as can be inferred from Figs.~\ref{pot_gc} and \ref{st_gc}. Here
the systematic error is about 6\%, while iterative gauge fixing might lead to
a discrepancy of $O(20\%)$.

\section{Color electric flux tube}

Studies of the pure $SU(2)$ gauge theory in the 
MAG~\cite{Haym93,Bali95,Komaflux}
have shown that the expectation values of the static color electric field and 
the monopole currents satisfy, to a good accuracy, the classical equations of 
motion and dual Amp\`ere's law, in agreement with the dual superconductor 
picture of confinement.

In this section we present first results of the microscopic structure of the
color electric flux tube in full QCD and in the pure $SU(3)$ gauge theory.
It is expected that long-range forces between quarks remain to exist in the
dynamical theory as well, because the color charge of quarks cannot be screened
locally by Higgs scalars (made out of gluons).

\subsection*{Observables}

We will primarily be concerned with local abelian operators 
\eq
\cO(s) =
\mbox{diag} (\cO_1(s),\cO_2(s),\cO_3(s)) \in U(1)\times U(1)\,.
\ee
For C-parity even operators, such as the action density and the monopole 
density, the correlator of $\cO(s)$ with the abelian Wilson loop $\cW_{\cC}$ is
given by~\cite{Bali:1994de} 
\begin{equation}
\langle  \cO(s) \rangle_{\cW} \equiv \frac{1}{3}
\frac{\langle\mbox{Tr}\,{\cO}(s) \mbox{Tr}\, \cW_\cC \rangle}{\langle
\mbox{Tr}\, \cW_\cC \rangle} - \frac{1}{3} \langle \mbox{Tr} {\mathcal
O}\rangle \,.\label{eq:operator_even}
\end{equation}
For C-parity odd operators $\cO$, such as the color electric field and the 
monopole current, we have
\begin{equation}
\langle \cO(s)  \rangle_{\cW}  \equiv \frac{\langle\mbox{Tr}\,({\cO}(s)\,
\cW_\cC)\rangle}{\langle\mbox{Tr}\, \cW_\cC \rangle}\,,
\label{eq:operator_odd}
\end{equation}
in analogy to the case of 
$SU(2)$~\cite{Haym93,Bali95,Zach:1995ni,Cheluvaraja:2002yj}. 

The action density $\rho_A^\cW$, the color electric field $E_i^\cW$, the 
monopole current $k^\cW$ and the
monopole density $\rho_M^\cW$, induced by the Wilson loop, are then given by
\begin{equation}
\rho_A^\cW(s) = \frac{\beta}{3} \sum_{\mu>\nu} \langle 
\mbox{diag} (\cos(\theta_1(s,\mu,\nu)),\cos(\theta_2(s,\mu,\nu)),
\cos(\theta_3(s,\mu,\nu)))\rangle_\cW \,,
\end{equation}
where $\theta_i(s,\mu,\nu) \equiv {\rm arg}(u_i(s,\mu,\nu))$ is the
plaquette angle,
\eq \label{eq:def-elf}
E_j^\cW(s) = {\rm i}\, \langle \mbox{diag} (\theta_1(s,4,j),
\theta_2(s,4,j),\theta_3(s,4,j))\rangle_\cW\,,
\ee
\eq \label{eq:def-cur}
k^\cW(^*s,\mu) = 2 \pi {\rm i}\, \langle \mbox{diag} 
(k_1(^*s,\mu),k_2(^*s,\mu),k_3(^*s,\mu)) \rangle_\cW\,, 
\ee
and
\begin{equation}\label{rhomon}
\rho_M^\cW(s) = 
\frac{1}{4}\sum_{\mu} \langle {\rm diag}\big(|k_1(^*s,\mu)|, 
|k_2(^*s,\mu)|, |k_3(^*s,\mu)|\big)\rangle_\cW\,,
\end{equation}
respectively. Out of the three `color' components of the observables only two 
are independent. In the following we shall take the average of the three
components.

As before, we take $\cC$ to be a loop of spatial extent $R$ and temporal 
extent $T$. The
four corners of the loop are placed at $(-R/2,0,0,0)$, $(R/2,0,0,0)$, 
$(-R/2,0,0,T)$ and $(R/2,0,0,T)$, and $s_4=T/2$ will be taken throughout this
section. 

\subsection*{Abelian flux tube}

Let us first consider the profile of the abelian flux tube. We take $R=10$, 
which on our lattices corresponds to a spatial separation of the static 
sources of $\approx 1$ fm, and $T=6$. We checked the $T$ dependence of part of
our results by comparing the numbers to $T=5$ and found only insignificant
changes, albeit for $R=6$, which justifies our choice. The spatial links are 
smeared as described in (\ref{smear}). 

\begin{figure}[tbp]
\begin{center}
\vspace{-1.75cm}
\epsfig{file=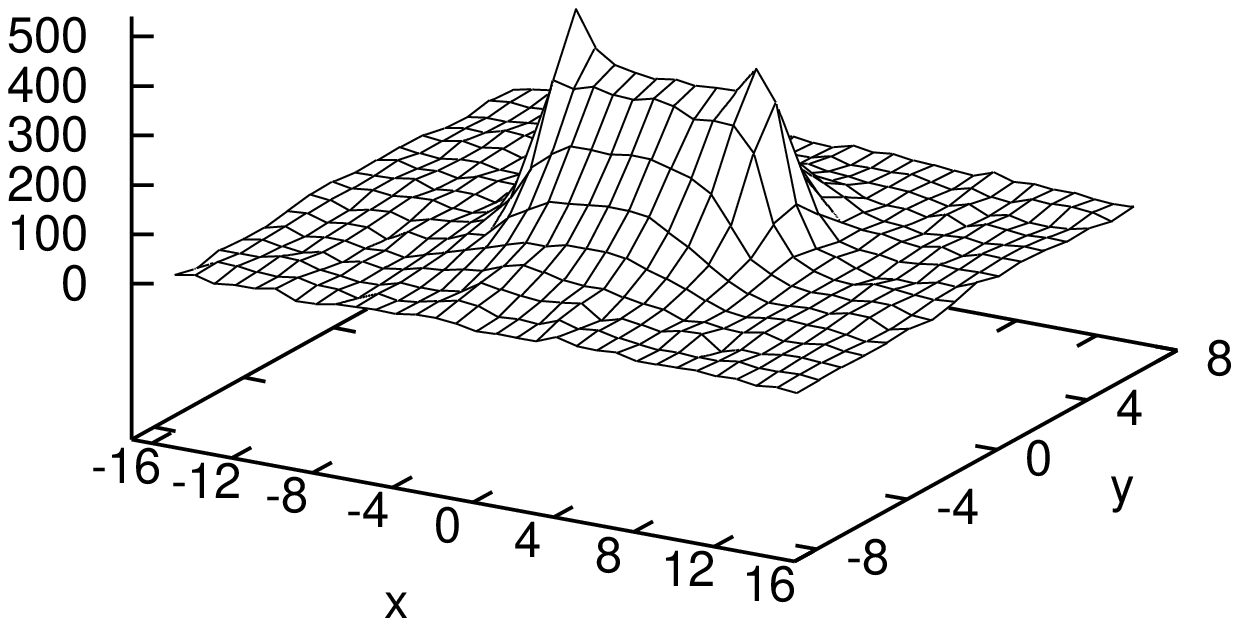,width=8cm,height=8cm,angle=270,clip=}\\[-6em]
\epsfig{file=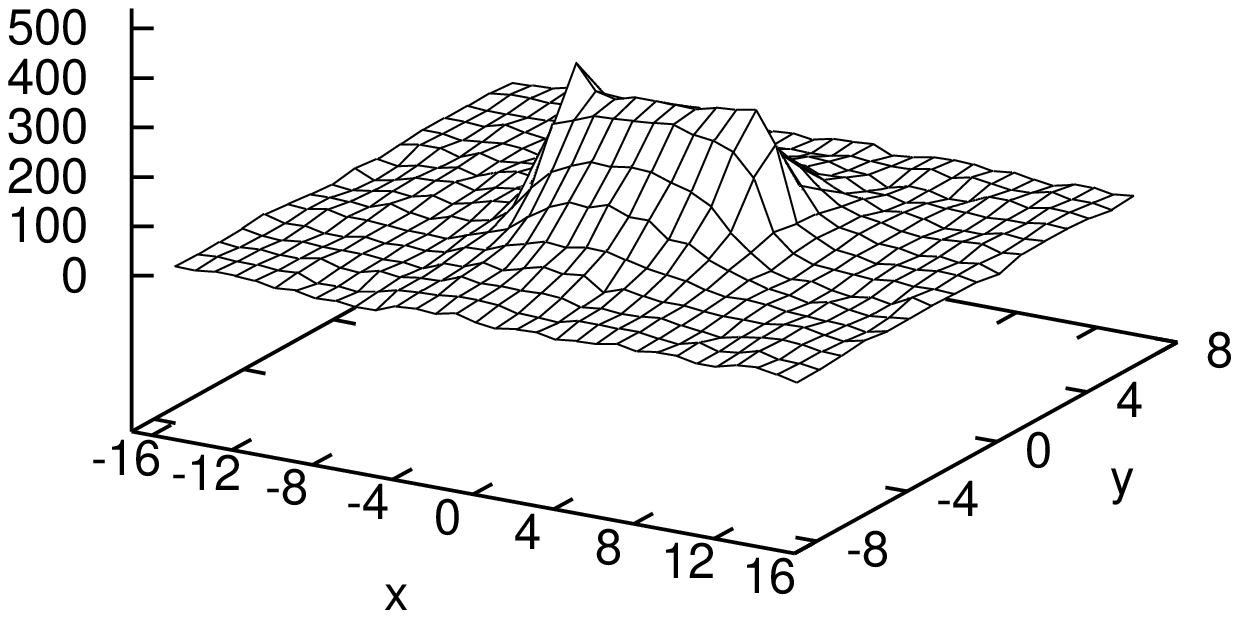,width=8cm,height=8cm,angle=270,clip=}
\end{center}
\caption{ The action density $\rho_A^\cW(s)\, r_0^4$ of the Abelian flux tube
as a function of $x=s_1$, $y=s_2$ at $z=s_3=0$ on the $16^3\, 32$ lattice in 
full (top) and quenched QCD (bottom) at $\beta=5.20$, $\kappa=0.1355$ and 
$\beta=6.0$, respectively.} \label{fig:actionFQ}
\end{figure}

\begin{figure}[hbp]
\begin{center}
\epsfig{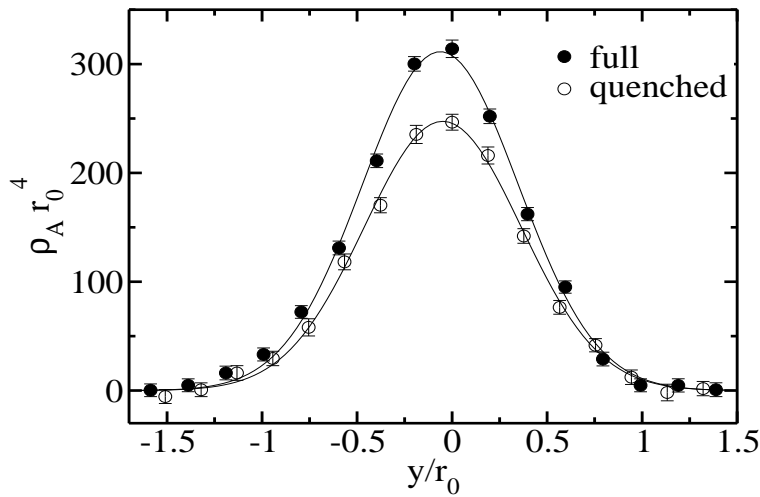}
\end{center}
\vspace{.25cm}
\caption{The action density $\rho_A^\cW(s)\, r_0^4$ of 
Fig.~\ref{fig:actionFQ} plotted across the flux tube at $x=0$.} 
\label{fig:AcFQ}
\end{figure}

In the following we shall also use the notation $x=s_1$, $y=s_2$ and $z=s_3$.
In Figs.~\ref{fig:actionFQ} and \ref{fig:AcFQ} we show the
action density $\rho_A^\cW(s)$ in full and quenched QCD, respectively. It 
appears that the action density in full QCD is
higher than in the quenched case, while their shapes are quite similar.

\begin{figure}[!t]
\vspace{.5cm}
\begin{center}
\includegraphics[width=7.9cm,height=5.4cm,angle=0]{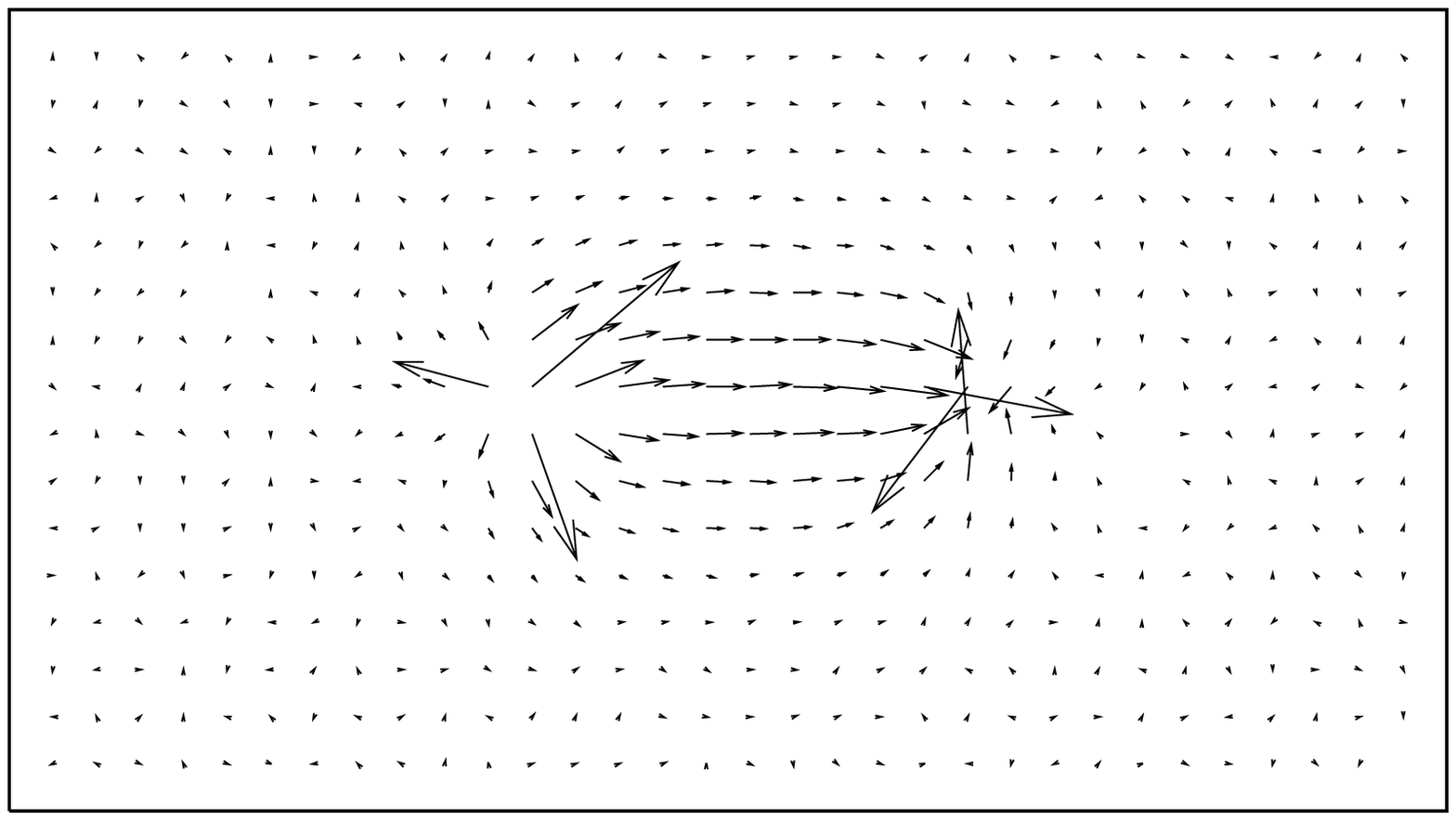}\\
\includegraphics[width=7.9cm,height=5cm,angle=0]{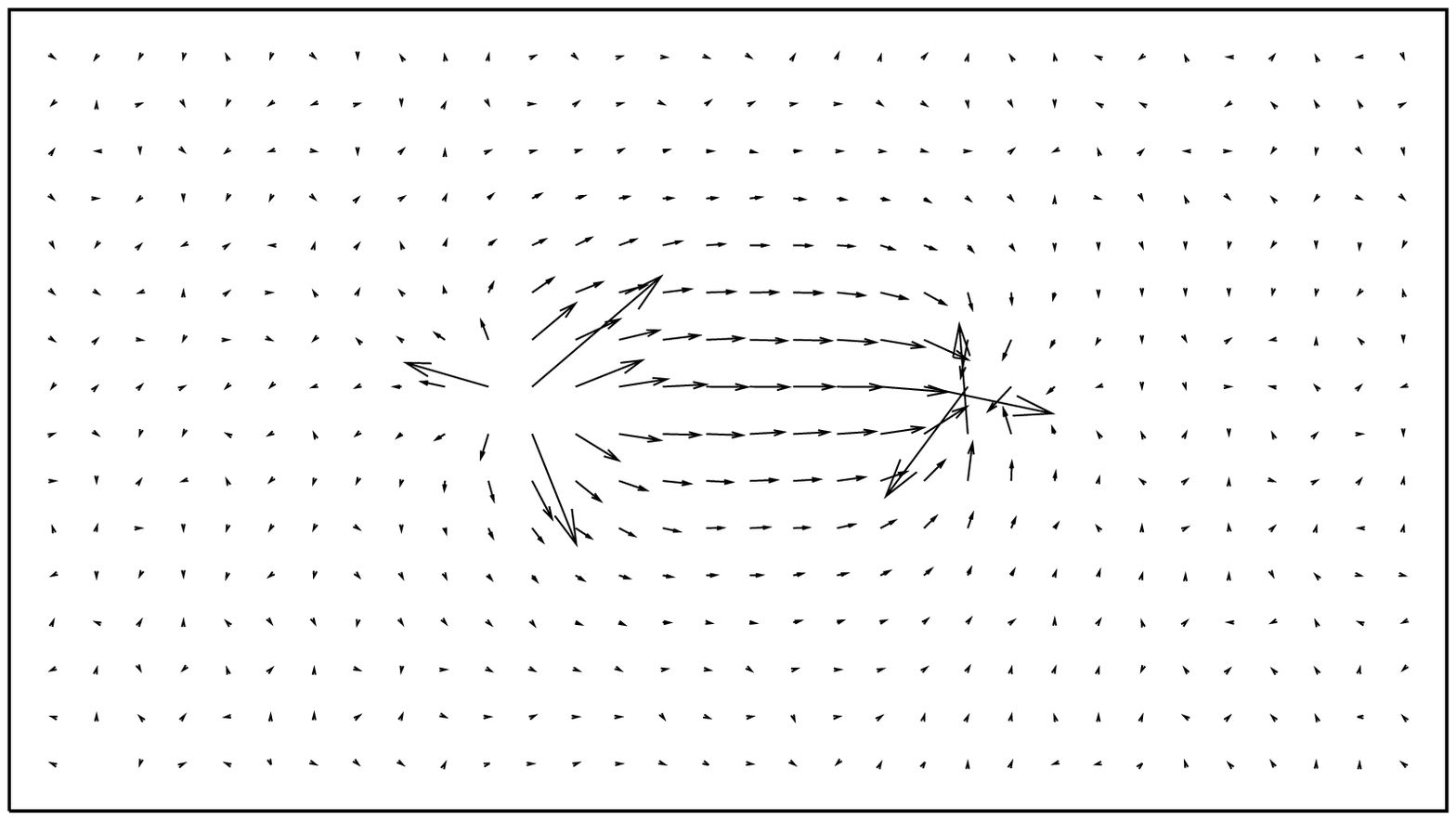}
\end{center}
\vspace{0.5cm}
\caption{Distribution of the color electric field $\vec{E}^\cW$ in full (top) 
and quenched 
QCD (bottom) in the ($x$,$y$) plane for the same lattices as in 
Fig.~\ref{fig:actionFQ}. The magnitude of $E^\cW$ is indicated by the length 
of the arrows.} \label{fig:EFQ}
\begin{center}
\vspace{1cm}
\epsfig{file=fig/EcFQ.eps,width=10cm,height=7cm,angle=0,clip=}
\end{center}
\vspace{0.25cm} 
\caption{The color electric field $E_1^\cW$ of Fig.~\ref{fig:EFQ} plotted 
across the flux tube at $x=0$.} \label{fig:EcFQ}
\end{figure}

We estimate the width $\delta$ of the abelian flux tube by fitting our data 
at $x=0$ to the function 
\begin{equation} \label{rhoAfit}
\rho_{A}(r_\perp) = \mbox{const.}\,
\exp\big(-(r_\perp-\epsilon)^2/\delta^2\big)\,,
\end{equation}
where $r_\perp$ is the distance of $s$ from the line connecting the static 
sources at $s_4=T/2$, and
$\epsilon$ is a displacement parameter of $O(a)$ accounting for 
a small shift of the true action density from its entry at $s$.
We obtain $\delta=0.29(1)$ fm, both in full and quenched QCD. This is a 
surprisingly small number, much smaller than any hadron radius. It tells us,
in particular, that already at interquark distances $\gtrsim 0.5$ fm the 
string model of hadrons becomes effective.

In Figs.~\ref{fig:EFQ} and \ref{fig:EcFQ} we show the distribution of the
color electric field $E_1^\cW$ in and around the flux tube. 
The electric field is purely longitudinal in a narrow region between the
sources of diameter $\approx$ six lattice spacings and practically zero 
outside. We fit $E_1^\cW$ at $x, z=0$ and for $y/r_0 > 0.5$ to an exponential:
\begin{equation}
E_1^\cW = \mbox{const.}\, \exp(-y/\lambda) \,.
\end{equation}
For the penetration length we find $\lambda = 0.15(1)$ fm in full QCD and
$\lambda = 0.17(1)$ fm in the quenched case. Whether the flux tube indeed 
narrows down in full QCD needs to be confirmed with higher statistics.

\begin{figure}[htbp]
\vspace{-1cm}
\begin{center}
\includegraphics[width=8.0cm,height=8cm,angle=270]{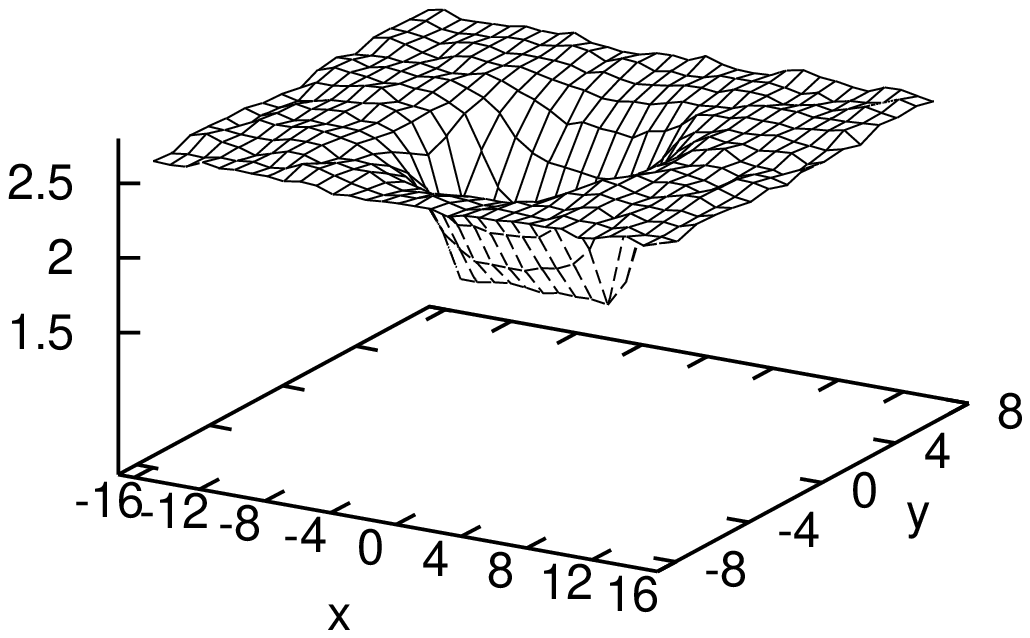}\\[-6em]
\includegraphics[width=8.0cm,height=8cm,angle=270]{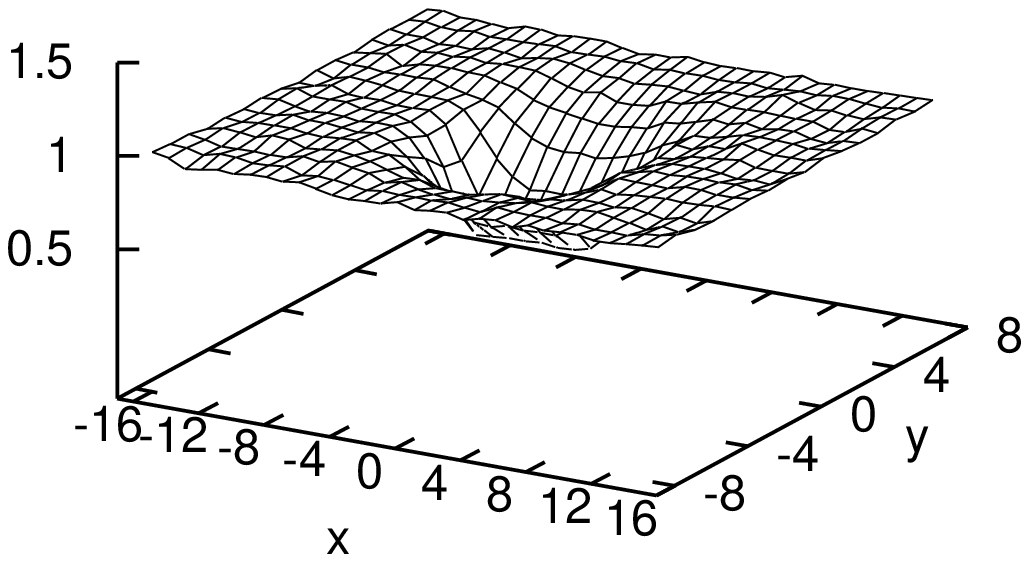}
\end{center}
\caption{The monopole density $\rho_M^\cW(s) r_0^3$ as a function of $x$, $y$ 
at $z=0$ in full (top) and quenched QCD (bottom) for the same lattices as in
Fig.~\ref{fig:actionFQ}.} \label{fig:monoFQ}
\end{figure}

\begin{figure}[htbp]
\vspace{0.5cm}
\begin{center}
\includegraphics[width=5.5cm,height=5.5cm, angle=0]{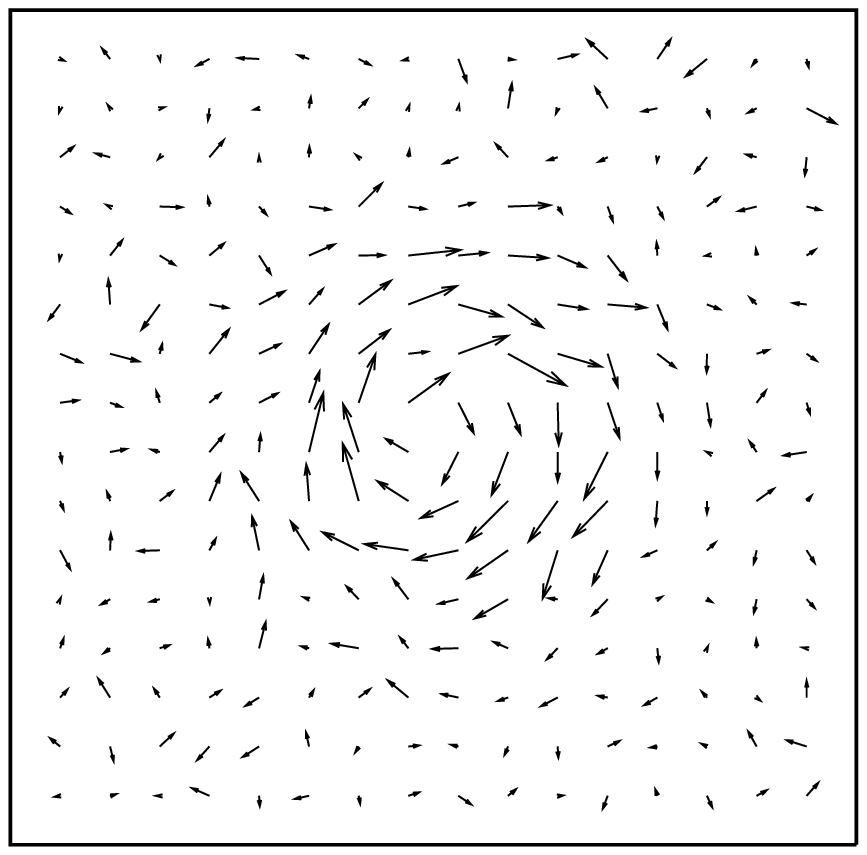}
\includegraphics[width=5.5cm,height=5.5cm, angle=0]{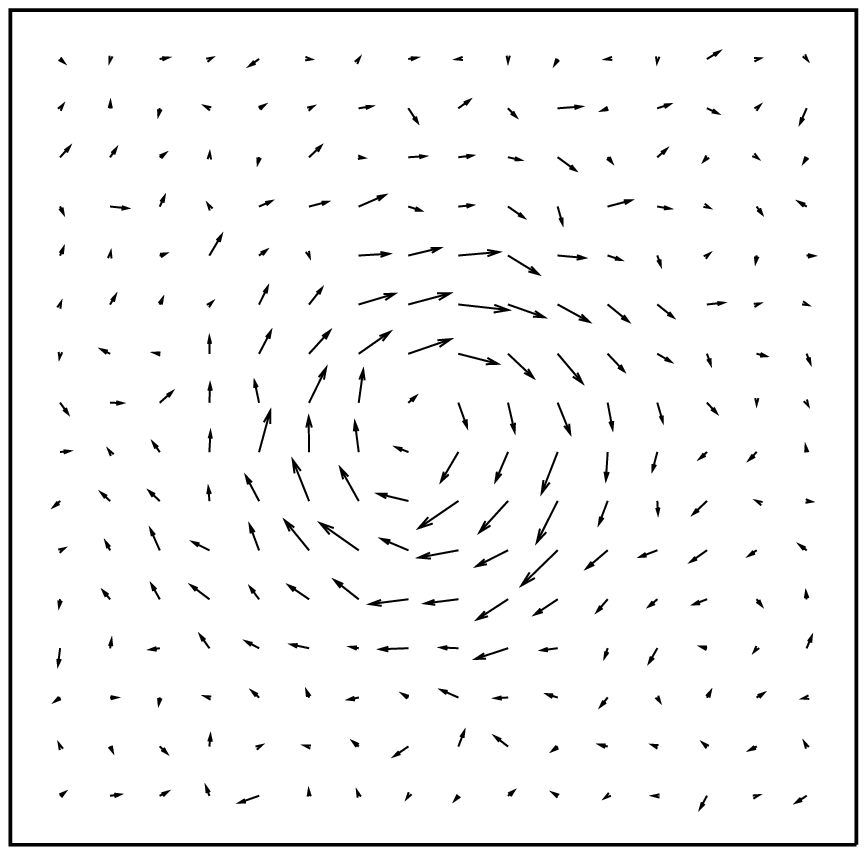}
\end{center}
\vspace{0.25cm}
\caption{The solenoidal monopole current $k^\cW r_0^3$ in the ($y$,$z$) plane
(i.e. perpendicular to the flux tube) at $x=0$ in full (left) and quenched 
QCD (right) for the same lattices as in Fig.~\ref{fig:actionFQ}.}
\label{fig:kFQ}
\end{figure}

\begin{figure}[htbp]
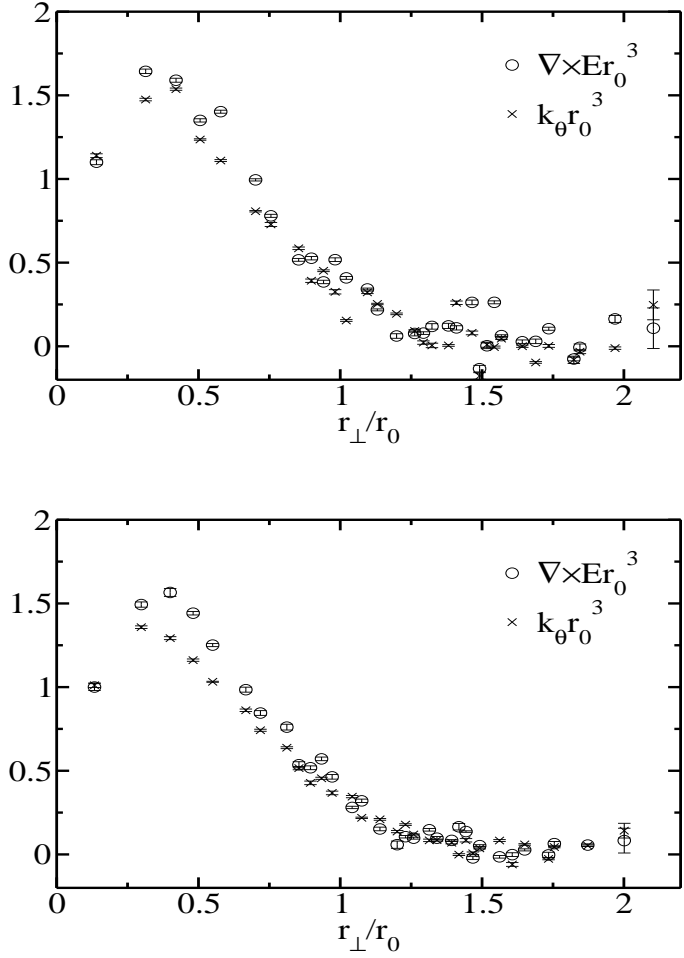

\vspace{0.5cm}
\begin{center}
\epsfig{file=fig/LondonFQ_1.eps,width=9cm,height=5.9cm,angle=0,clip=}\\[2em]
\epsfig{file=fig/LondonFQ_2.eps,width=9cm,height=5.9cm,angle=0,clip=}
\end{center}
\vspace{0.5cm} \caption{Test of dual Amp\`ere's law in full (top) and 
quenched QCD (bottom) for the same lattices as in Fig.~\ref{fig:actionFQ}.} 
\label{fig:LondonFQ}
\vspace{0.5cm} 
\end{figure}

In Figs.~\ref{fig:monoFQ} we show the monopole density $\rho_M^\cW(s)$. 
We see again that outside the flux tube the monopole density is more than two 
times larger in full QCD than in the quenched case. Inside the flux tube the
monopole density is strongly suppressed. This indicates that the expectation
value of the dual Higgs field vanishes inside the flux tube, in agreement with
the dual superconductor model of the vacuum. In this model we furthermore
expect that the 
monopole currents forms a solenoidal (i.e. azimuthal) supercurrent which 
constricts the color electric field into flux tubes, thereby satisfying 
the dual Amp\`ere law:
\begin{equation}
\vec{k} = \vec{\nabla} \times \vec{E}^\cW \label{ampere} \,.
\end{equation}
In Fig.~\ref{fig:kFQ} we show the transverse components of the
monopole current at $x=0$ in the ($y$,$z$) plane (i.e. perpendicular to the 
flux tube), and in Fig.~\ref{fig:LondonFQ} we compare the l.h.s and r.h.s. of
eq.~(\ref{ampere}). In the latter figure $R=6$ was used in order to reduce the
errors. We find that the dual Amp\`ere law is approximately satisfied in 
both full and quenched QCD.  
So far this has only been verified in the pure $SU(2)$ gauge 
theory~\cite{Haym93,Bali95}.


\subsection*{Monopoles versus photons}

Further insight into the confinement mechanism can be obtained by probing the 
flux tube in terms of the monopole and photon part of the abelian 
gauge field separately. To do so, we simply have to replace the lattice 
abelian gauge field $\theta_i(s,\mu)$ in the various probes by  
$\theta_i^{\rm mon}(s,\mu)$ and $\theta_i^{\rm ph}(s,\mu)$, respectively.
We have done calculations in full QCD (here on the $24^3\,48$ lattice at
$\beta=5.29$, $\kappa=0.1355$) and in the quenched theory. To enhance the 
signal, the monopole and photon fields are smeared as before. Qualitatively,
we find no difference between full QCD and the quenched theory.

\begin{figure}[tbp]
\vspace{-1cm}
\begin{center}
\epsfig{file=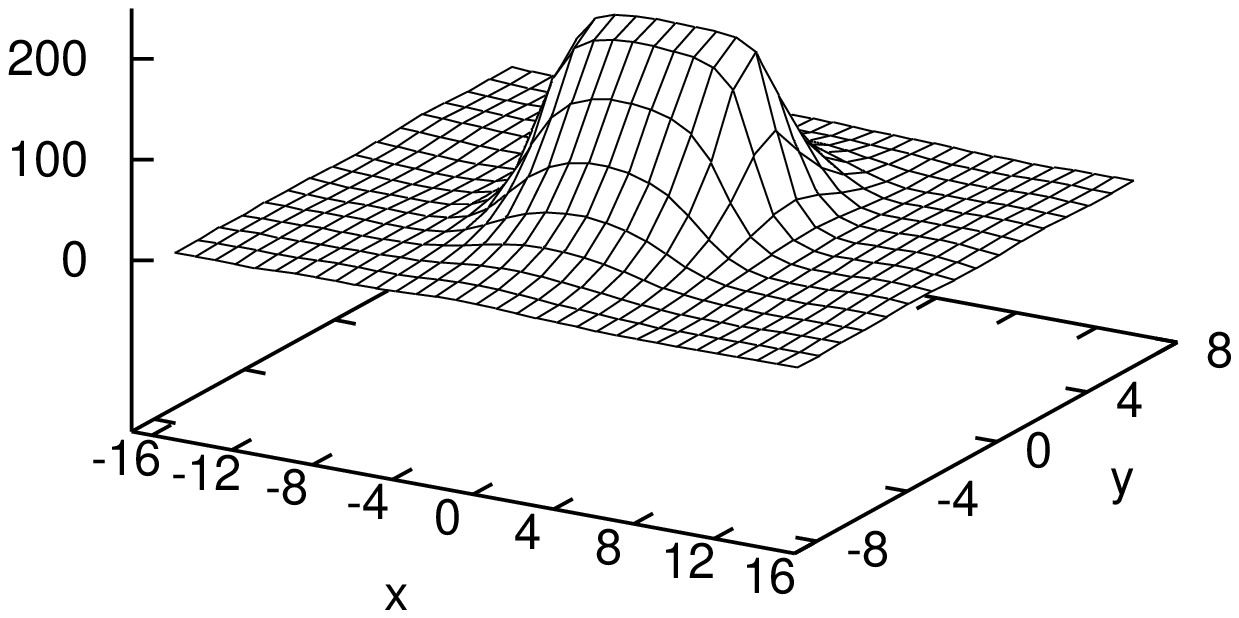,width=6cm,angle=270,clip=}\\[-4em]
\epsfig{file=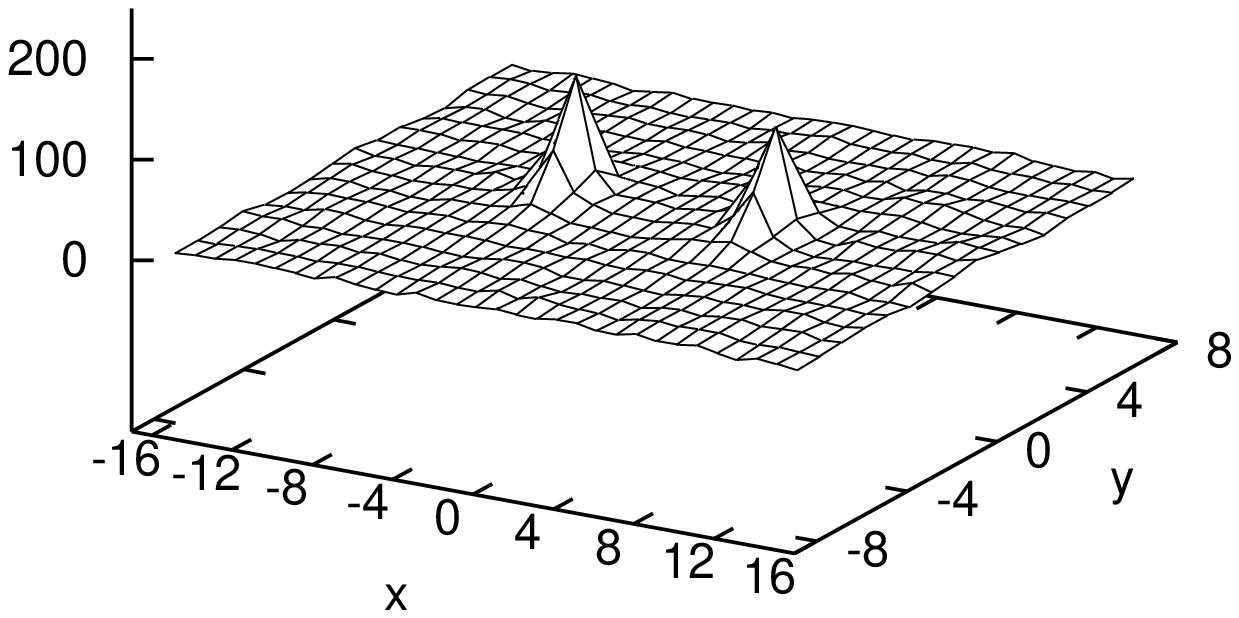,width=6cm,angle=270,clip=}
\end{center}
\caption{Monopole (top) and photon part (bottom) of the action density
$\rho_A^\cW r_0^4$ on the (quenched) $16^3\,32$ lattice at $\beta=6.0$.} 
\label{fig:actionMP}
\end{figure}
\vspace{0.5cm}
\begin{figure}
\begin{center}
\vspace{0.25cm}
\includegraphics[width=8.5cm,height=4.5cm, angle=0]{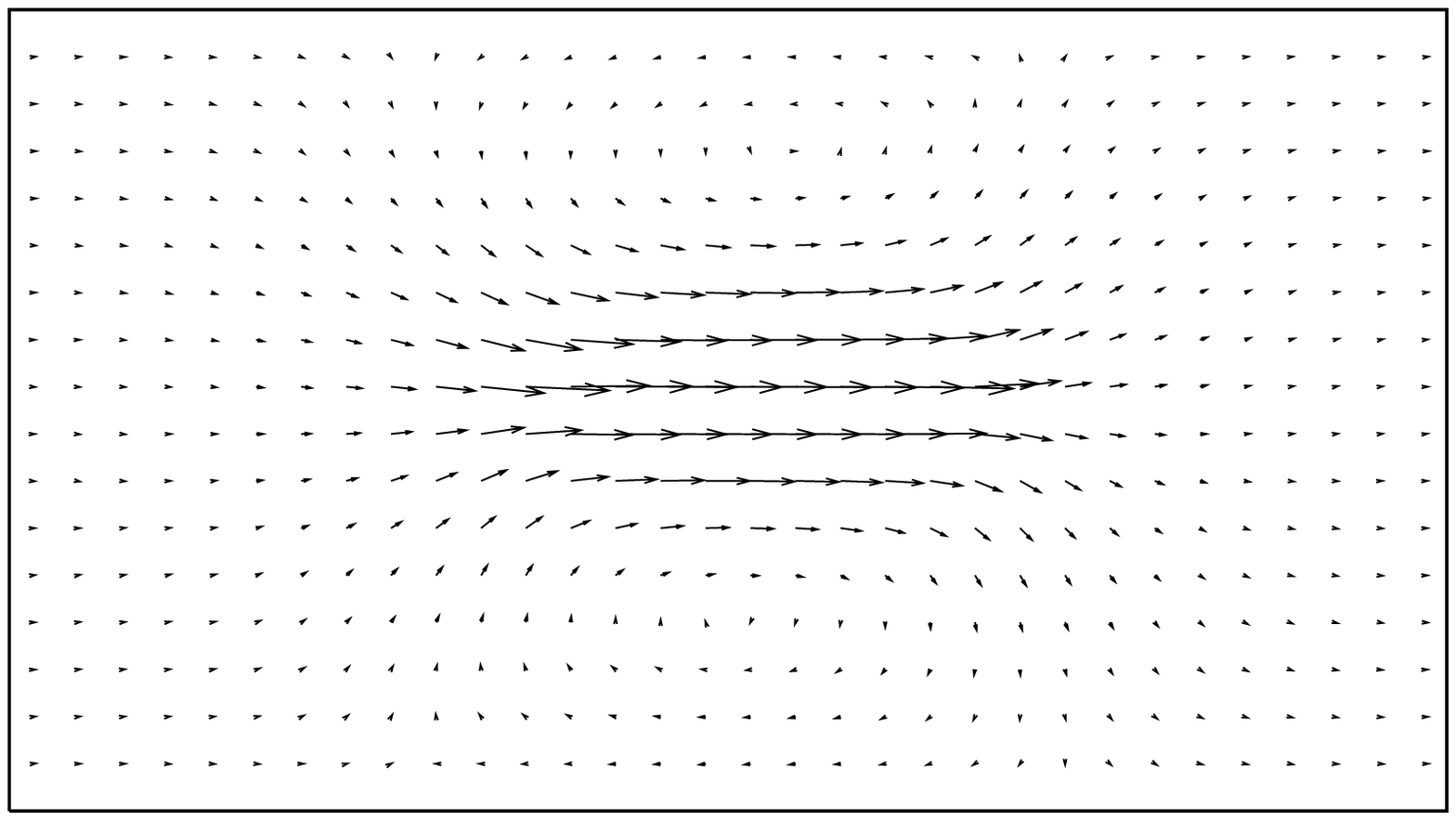}
\includegraphics[width=8.5cm,height=4.5cm, angle=0]{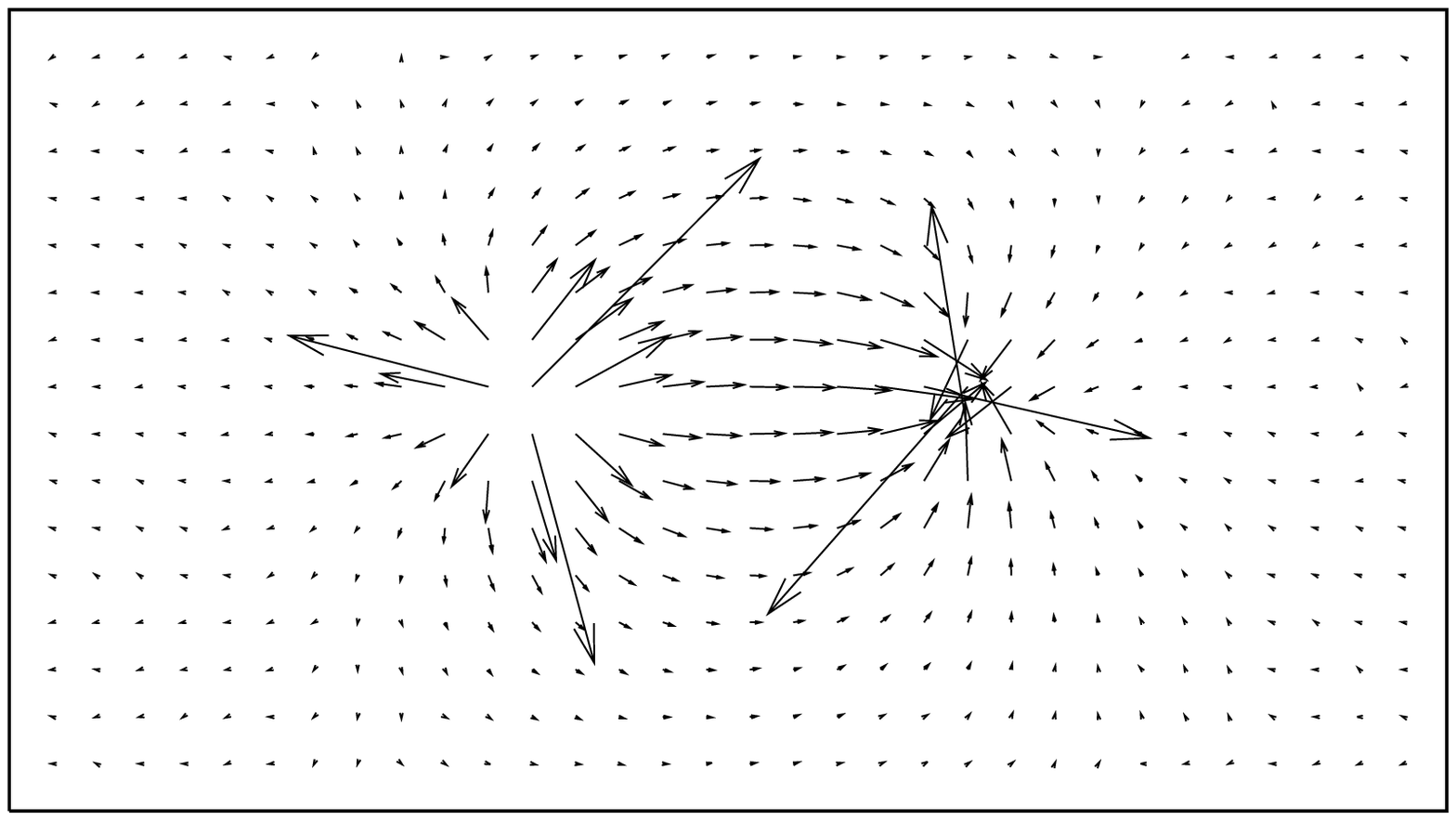}
\end{center}
\caption{Distribution of the monopole (top) and photon part of the
color electric field $\vec{E}^\cW$ (bottom) on the $16^3\,32$ lattice at 
$\beta=6.0$. For better visibility the monopole part is enhanced by a factor 
of two relative to the photon part.} \label{fig:EMP}
\end{figure}

\begin{figure}[tbp]
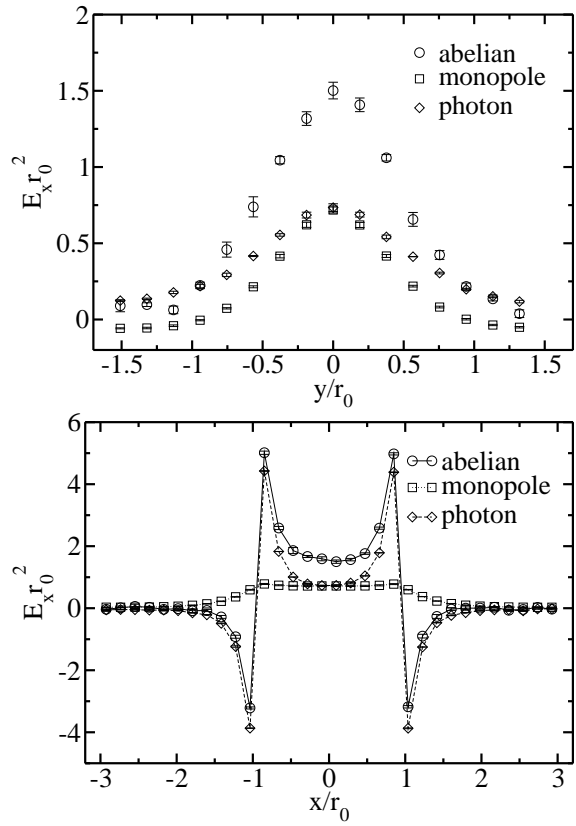

\vspace{.5cm}
\begin{center}
\epsfysize=15.0cm \epsfxsize=8.cm
\epsfig{file=fig/EcpMP_1.eps,width=7.5cm,angle=0,clip=}
\epsfig{file=fig/EcpMP_2.eps,width=7.5cm,angle=0,clip=}
\end{center}
\caption{Monopole and photon part of the color electric field $E_1^\cW$
in the ($y$,$z$) plane at $x = 0$ (top) and parallel to the flux tube 
(bottom).} \label{fig:EcpMP}
\vspace*{0.25cm}
\end{figure}

In Fig.~\ref{fig:actionMP} we show the action density
$\rho_A^\cW$ of the monopole and photon part of the gauge field. We see that 
the action density originates almost entirely from the monopole part, while the
photon contributes a Coulomb field around the static charges only. The width
of the flux tube is unchanged: $\delta = 0.29(1)$ fm as before.
In Figs.~\ref{fig:EMP} and \ref{fig:EcpMP} we show the distribution of the
color electric field. We see that the monopole part of the field has no 
sources. The sources show up in the photon part only. We furthermore see that
outside the flux tube the monopole and photon parts of the electric field
largely cancel, while they add in the interior of the tube. We have attempted
to fit the photon field by a Coulomb ansatz. While the transverse component
could be well fitted, we failed for the longitudinal component (i.e. parallel 
to the flux tube).

\subsection*{String breaking}

In the presence of dynamical quarks we expect that the flux tube (string) 
breaks if the static charges are separated far enough. It has been estimated
that this will happen at a distance of
$\approx 1.2$ fm for our quark masses of $m_q \approx 100$ 
Mev~\cite{allt,Bali:2000vr}. This does not mean that the string state
vanishes from the spectrum of the transfer matrix. It only ceases to be the
state of lowest energy in the corresponding channel. In QCD string breaking 
has so far only been observed at
finite temperature close to the deconfining phase transition~\cite{PT}, but
never at zero temperature~\cite{allt,SESA96,CPPA99,Schi00}. The recent 
finding~\cite{Forcrand} of string breaking from Wilson loops in the case
of adjoint quarks in three-dimensional $SU(2)$ gauge theory indicates that
that this should be possible at zero temperature too.
Though our prime motivation for the following
investigation was to detect string
breaking and shed some light on the dynamics that drives it, we like to 
stress that the unbroken string is of quite some interest as well from the 
point of view of the confinement problem.

\begin{figure}[!t]
\begin{center}
\vspace{-0cm}
\includegraphics[width=11cm,height=6cm,angle=0]{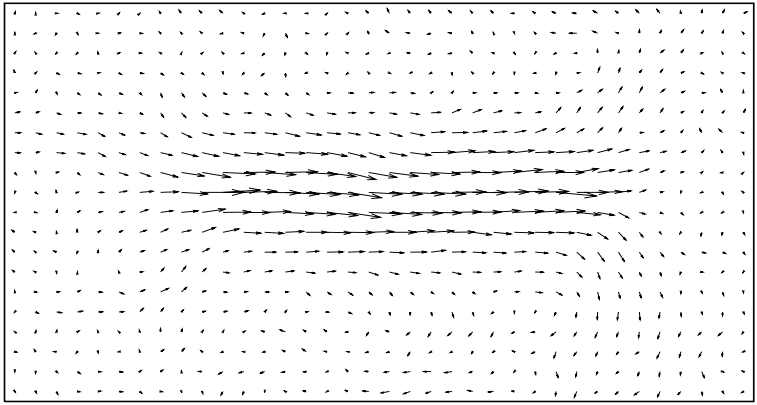}
\end{center}
\vspace{0.5cm} \caption{The monopole part of the color electric field 
$\vec{E}^\cW$ on the $24^3\,48$ lattice. Only the region $-18 \le x \le 17$,
$-10 \le y \le 9$ is shown.} \label{fig:Efm1018}
\end{figure}

The calculations in this section are done on the $24^3\,48$ lattice at
$\beta=5.29$, $\kappa=0.1355$. On this lattice  $m_\pi/m_\rho \approx 0.7$.
The difference to our previous calculation is that we consider large Wilson 
loops. We take $R=18$, which corresponds to a separation of the static charges
of $\approx 1.6$ fm, and $T=10$. It is important to choose $T$ large as well
in order to increase the chance of string breaking~\cite{Trot99}. In 
Fig.~\ref{fig:Efm1018} we show the monopole part of the color electric field
$\vec{E}^\cW$. The restriction to the monopole part allowed us to obtain a
clean signal even at $x=0$. We do not observe any sign of string breaking.
Furthermore, the flux tube does not show any broadening effect.
A fit of the form (\ref{rhoAfit}) gives $\delta=0.30(3)$ fm at $x=0$. The same
value was found for $R=10$. If the long-range properties of QCD were described
by the Nambu-Goto effective string action, we would have expected the 
transverse extension
of the flux tube to increase logarithmically with $R$~\cite{Lusc81}. A
similar observation has been made in the pure $SU(2)$ gauge 
theory~\cite{bali3}.
Perhaps the abelian flux tube is described by the Ramond effective string 
action instead, which does not give rise to any broadening 
effect~\cite{Oles85}.

It is perhaps not surprising that we do not observe any string breaking 
(yet)~\cite{MILC01}. String breaking is expected to occur if
$\exp(-2E_{sl}(R+T)) > \exp(-\sigma RT)$, where $E_{sl}$ is the binding energy
of the static-light meson. This is only the case if $T \gtrsim 3$ 
fm~\cite{CPPA99}.

\section{Effective monopole action}

We have seen that the vacuum undergoes several changes if dynamical color 
electric charges are introduced. We shall study now how this will affect the 
effective monopole action.

There are three types of monopole currents (\ref{current}), of which two are
independent. For simplicity we take into account only one of them, thus
integrating out the other two~\cite{aekms}. For the time being, we assume the
form of the effective monopole action in full QCD to be the same as in the
quenched theory~\cite{Yamagishi}. It is composed of 27 types of two-point
interactions, one four-point interaction and one six-point interaction:
\begin{equation}
S(k)=\sum_{i=1}^{29} G_i S_i (k)\,,
\end{equation}
where $G_i$ are the coupling constants which need to be determined. This we
will do by employing an extended Swendsen method~\cite{Shiba:1995pu}. In this 
section we shall write $k(s,\mu)$ instead of $k(^*s,\mu)$ for the sake of 
simplicity. Explicitly, we then have:
\vspace*{0.25cm}

\noindent
{\it Two-point interaction for parallel currents}

\begin{equation}
S_{i}(k)=\sum_{s}\sum_{\mu =1}^{4}  k(s,\mu) \, k_{i}(s,\mu)\,, \quad 
i=1, \cdots, 25\,,
\label{mocu}
\end{equation}
where the $k_{i}(s,\mu)$ are given in Table \ref{tabmon}.
\vspace*{0.25cm}

\begin{table}[t]
\begin{center}
\begin{tabular}{|c|l||c|l|}
\hline
$i$ & $k_{i}(s,\mu)$  &$i$ & $k_{i}(s,\mu)$    \\
\hline $ 1 $&$  k(s,\mu)                                              $ &
$ 14$&$  k(s+2\hat{\mu}+\hat{\nu}+\hat{\rho},\mu)              $\\
$ 2 $&$  k(s+\hat{\mu},\mu)                                    $  &
$ 15$&$  k(s+\hat{\mu}+2\hat{\nu}+\hat{\rho},\mu)              $\\
$ 3 $&$  k(s+\hat{\nu},\mu)                                    $&
$ 16$&$  k(s+2\hat{\nu}+\hat{\rho}+\hat{\sigma},\mu)           $\\
$ 4 $&$  k(s+\hat{\mu}+\hat{\nu},\mu)                          $&
$ 17$&$  k(s+2\hat{\mu}+\hat{\nu}+\hat{\rho}+\hat{\sigma},\mu) $\\
$ 5 $&$  k(s+\hat{\nu}+\hat{\rho},\mu)                         $&
$ 18$&$  k(s+\hat{\mu}+2\hat{\nu}+\hat{\rho}+\hat{\sigma},\mu) $\\
$ 6 $&$  k(s+2\hat{\mu},\mu)                                   $ &
$ 19$&$  k(s+2\hat{\mu}+2\hat{\nu},\mu)                        $\\
$ 7 $&$  k(s+2\hat{\nu},\mu)                                   $ &
$ 20$&$  k(s+2\hat{\nu}+2\hat{\rho},\mu)                       $\\
$ 8 $&$  k(s+\hat{\mu}+\hat{\nu}+\hat{\rho}+\hat{\sigma},\mu)  $&
$ 21$&$  k(s+3\hat{\mu},\mu)                                   $\\
$ 9 $&$  k(s+\hat{\mu}+\hat{\nu}+\hat{\rho},\mu)               $&
$ 22$&$  k(s+3\hat{\nu},\mu)                                   $\\
$ 10$&$  k(s+\hat{\nu}+\hat{\rho}+\hat{\sigma},\mu)            $&
$ 23$&$  k(s+2\hat{\mu}+2\hat{\nu}+\hat{\rho},\mu)             $\\
$ 11$&$  k(s+2\hat{\mu}+\hat{\nu},\mu)                         $&
$ 24$&$  k(s+\hat{\mu}+2\hat{\nu}+2\hat{\rho},\mu)             $\\
$ 12$&$  k(s+\hat{\mu}+2\hat{\nu},\mu)                         $&
$ 25$&$  k(s+2\hat{\nu}+2\hat{\rho}+\hat{\sigma},\mu)          $\\
$ 13$&$  k(s+2\hat{\nu}+\hat{\rho},\mu)                        $& & \\
\hline
\end{tabular}
\end{center}
\vspace*{0.5cm}
\caption{The monopole currents $k_{i}(s,\mu)$ entering eq.~(\ref{mocu}).} 
\label{tabmon}
\end{table}

\noindent
{\it Two-point interaction for orthogonal currents} 

\begin{equation}
S_{26}(k)=\sum_{s}\sum_{\mu \ne \nu } k(s,\mu) \, k(s-\hat{\mu}-2\hat{\nu},
\nu)\,, \\
\end{equation}
\begin{equation}
S_{27}(k)=\sum_{s}\sum_{\mu \ne \nu \ne \sigma}  k(s,\mu) \,  k(s-\hat{\mu}-2\hat{\nu}-2\hat{\sigma},\sigma)\,.
\end{equation}

\noindent
{\it Four-point interaction}
\begin{equation}
S_{28} (k) = \sum_s \left( \sum_{\mu=-4}^4k(s,\mu)^2 \right)^2.
\end{equation}

\noindent
{\it Six-point interaction}
\begin{equation}
S_{29} (k) = \sum_{s} \left( \sum_{\mu=-4}^4 k(s,\mu)^2 \right)^3.
\end{equation}

The calculations are done on the $24^3\,48$ lattices listed in Table~1, both 
in full and quenched QCD. After fixing the gauge fields to the MAG, we employ 
a type-II block spin transformation~\cite{Ivanenko} with up to $n=4$ blocking 
steps. The final outcome is an action at the physical length scale $b=na$. We 
believe that the monopole action is effective at scales $0.4 \lesssim b 
\lesssim 0.8$ fm.

In Fig.~\ref{fig:g1-vs-b} we show the self-coupling $G_1$ as a function of 
$b$. We see that in full QCD $G_1$ 
is systematically smaller than in the quenched theory for all values of $b$.
In Fig.~\ref{pi-rho} we plot the self-coupling $G_1$ as a function of 
$m_{\pi}/m_{\rho}$ for our smallest $b$ value. We find that $G_1$ decreases
with decreasing quark mass.

\begin{figure}[tbp]
\begin{center}
\epsfig{file=fig/g1-vs-bx.eps,width=10cm,height=9cm,clip=}
\end{center}
\caption{The monopole self-coupling $G_1$, as
a function of the physical length scale $b$. The symbols are: $n=1$ $(${\Large
$\bullet$}$)$, $n=2$ $(\blacksquare)$, $n=3$ $(\blacklozenge)$, $n=4$
$(\blacktriangle)$ in full QCD, and $n=1$ $(${\Large $\circ$}$)$, $n=2$
$(\square)$, $n=3$ $(\lozenge)$, $n=4$ $(\triangle)$ in quenched QCD.}
\label{fig:g1-vs-b}
\begin{center}
\vspace{0.5cm}
\epsfig{file=fig/pirho09.eps,width=10cm,height=9cm,clip=}
\end{center}
\caption{The dependence of $G_1$ on the ratio $m_{\pi}/m_{\rho}$ for $b=0.09$ 
fm. The entry at $m_\pi/m_\rho = 1$ corresponds to the quenched theory.} 
\label{pi-rho}
\end{figure}

\begin{figure}[htbp]
\begin{center}
\epsfysize=9.cm \epsfxsize=10cm \epsfbox{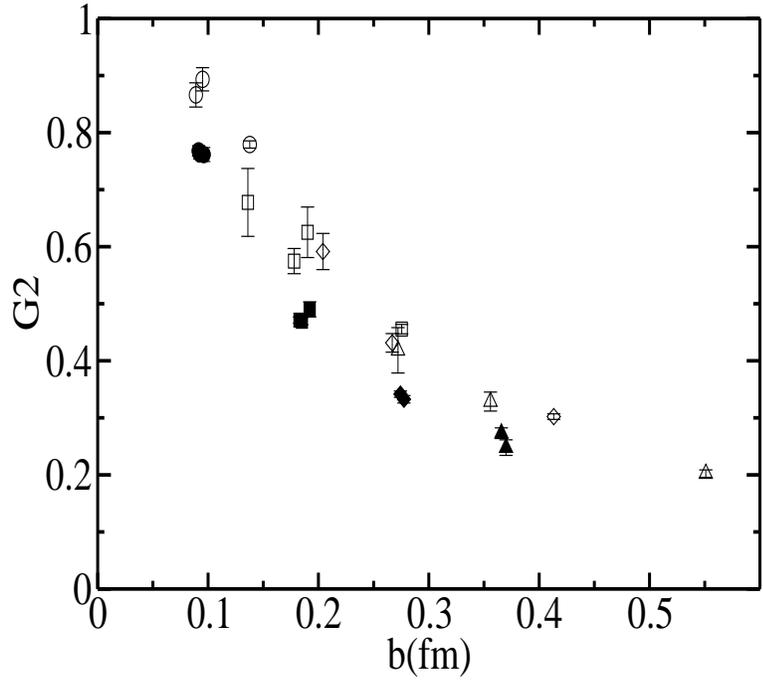} \\
\vspace{1.5cm}
\epsfysize=9.cm \epsfxsize=10cm \epsfbox{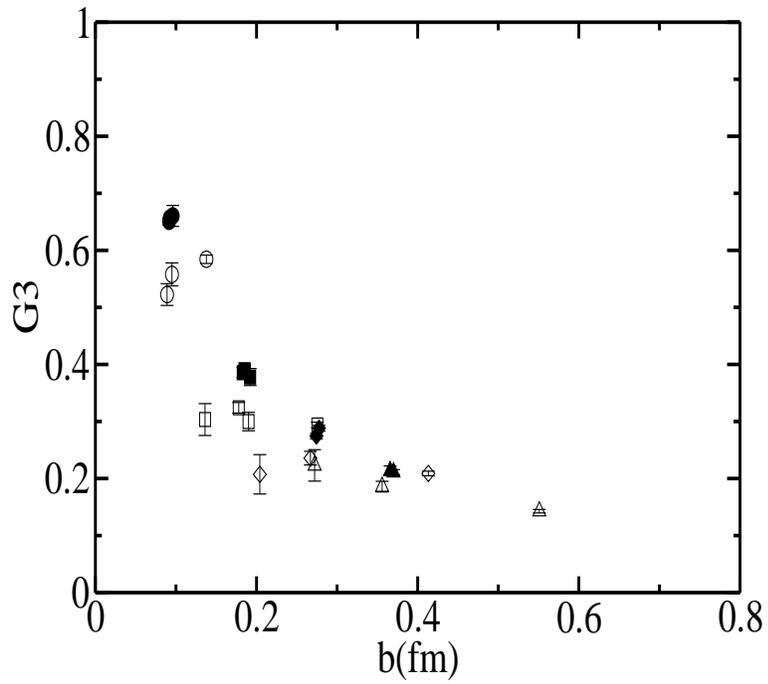}
\end{center}
\caption{The coupling constants $G_2$ and $G_3$ as a function of $b$. The
symbols are as in Fig.~\ref{fig:g1-vs-b}.} \label{fig:g2-g3-vs-b}
\end{figure}

\begin{figure}[htbp]
\begin{center}
\epsfysize=9cm \epsfxsize=10cm \epsfbox{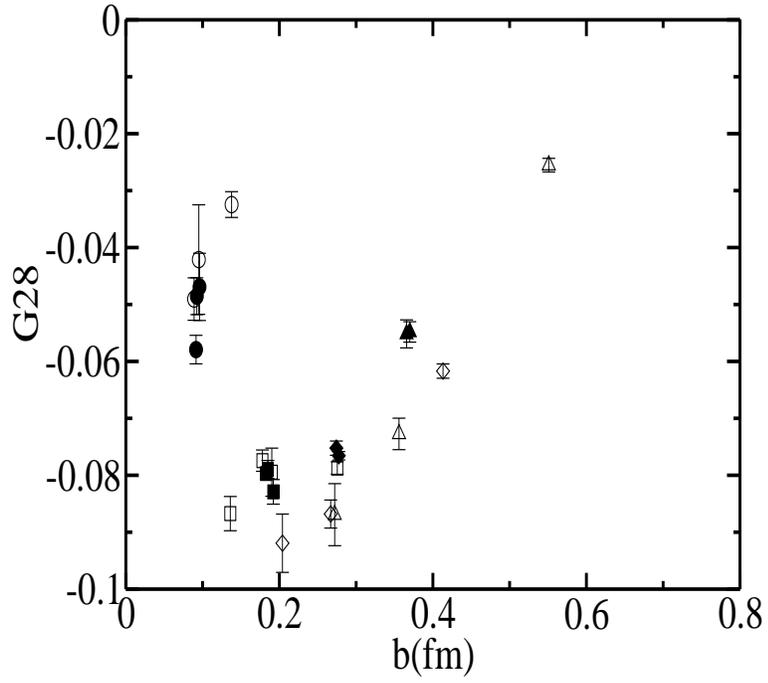} \\
\vspace{1.5cm}
\epsfysize=9cm \epsfxsize=10cm \epsfbox{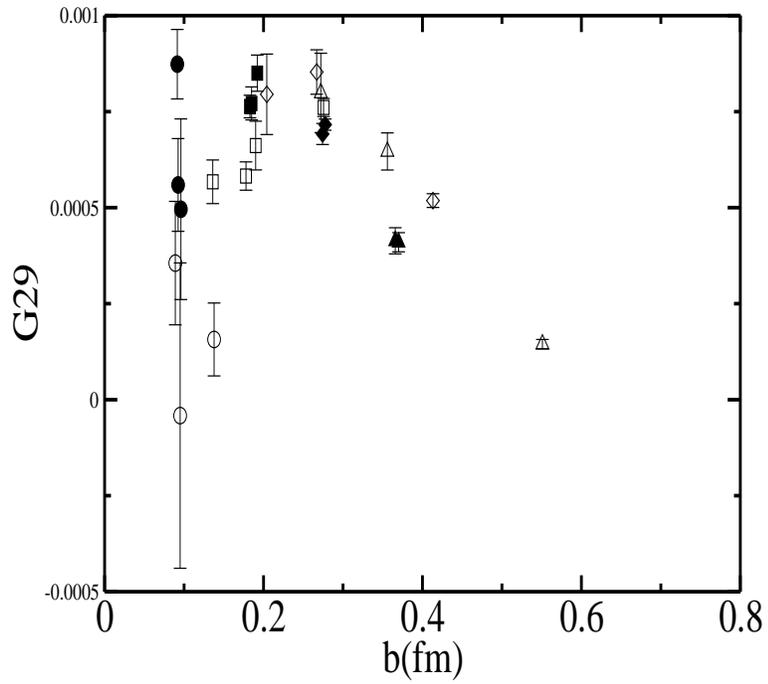}
\end{center}
\caption{The coupling constants $G_{28}$ and $G_{29}$ of the four-point and
six-point interactions, respectively, as a function of $b$. The symbols are as
in Fig.~\ref{fig:g1-vs-b}.} \label{fig:g28-g29-vs-b}
\end{figure}

\begin{figure}[htbp]
\begin{center}
\epsfysize=9cm \epsfxsize=10cm \vspace{0.1cm} \epsfbox{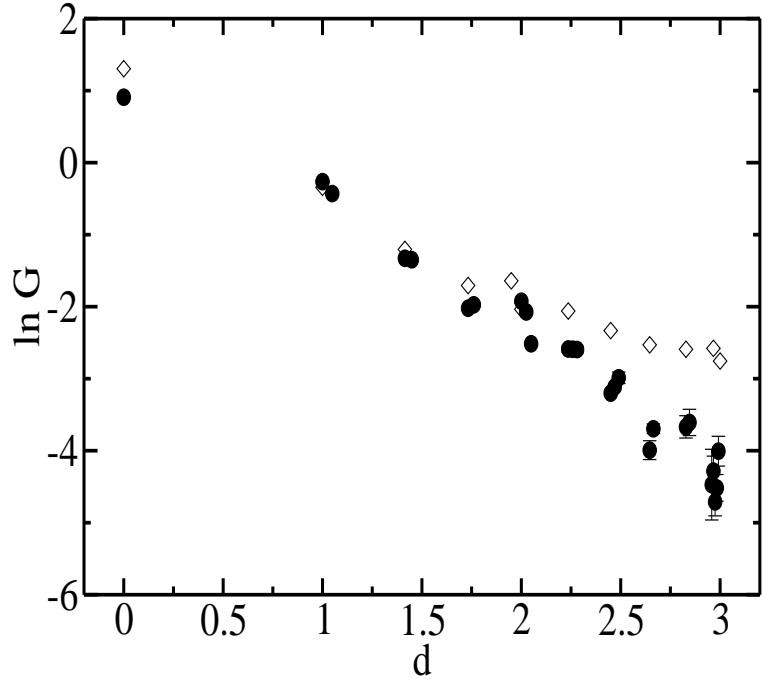} \\
\vspace{1.4cm}
\epsfysize=9cm \epsfxsize=10cm \epsfbox{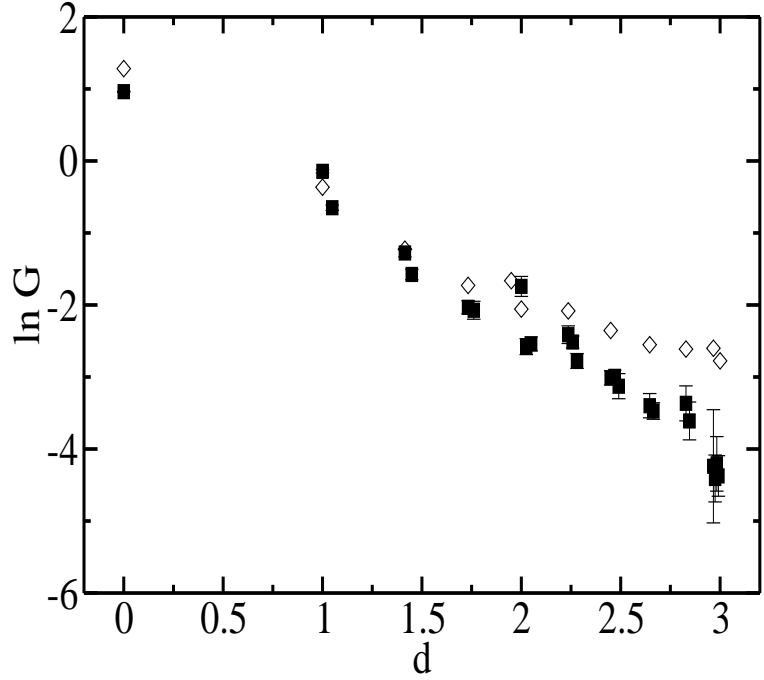}
\end{center}
\caption{The two-point coupling $G$ of the monopole currents $k(s,\mu)$ 
and $k(s^\prime,\mu)$ as a function of distance
$d=\sqrt{\sum_\mu(s_\mu-s^\prime_\mu)^2}$ in full $(${\Large $\bullet$}$)$
and quenched QCD $(\blacksquare)$, compared to the Coulomb propagator 
$(\lozenge)$.} \label{fig:comp}
\end{figure}

\begin{figure}[htbp]
\begin{center}
\epsfysize=9cm \epsfxsize=10cm \epsfbox{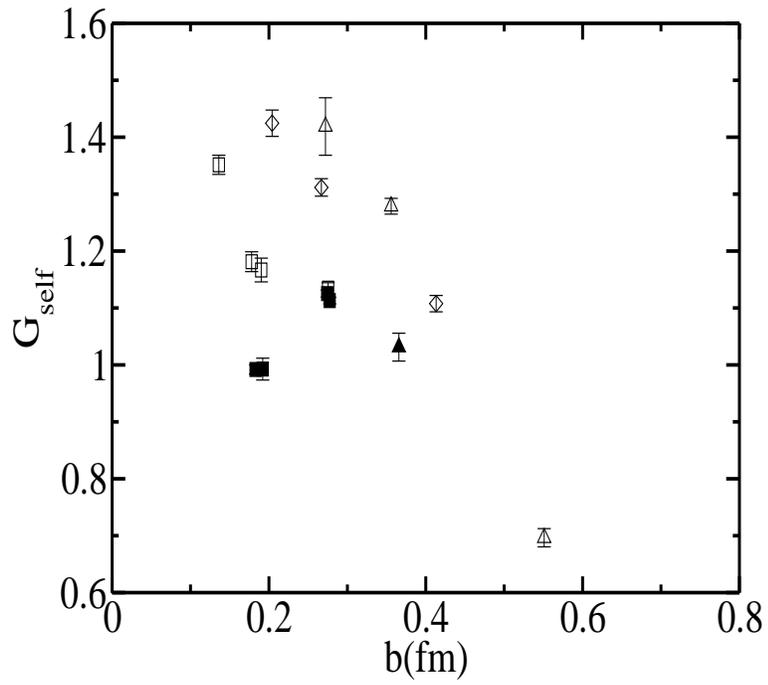} \\
\vspace{1.4cm} 
\epsfysize=9cm \epsfxsize=10cm \epsfbox{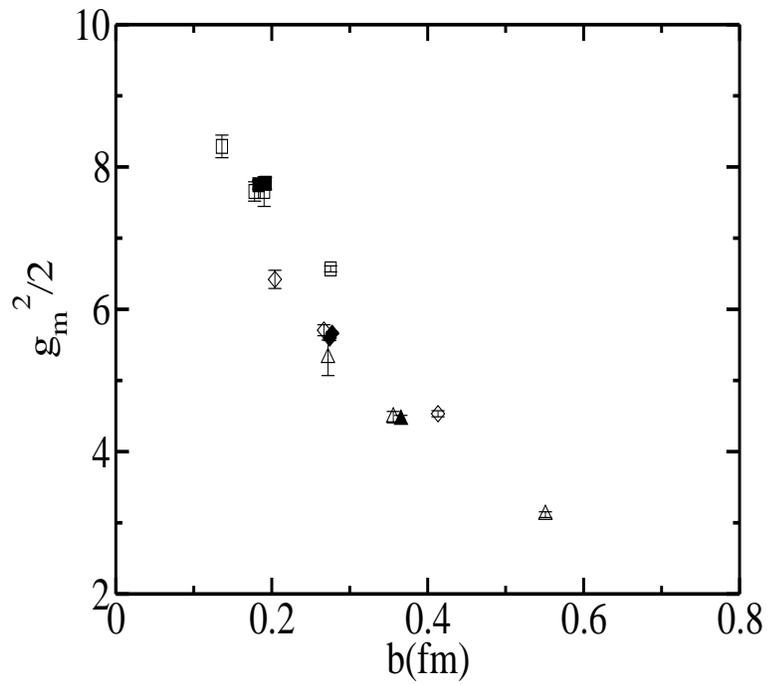}
\end{center}
\caption{The self-coupling $G_{\rm self}$ and the Coulomb coupling $g_{\rm m}$
as a function of $b$. The symbols are as in Fig.~\ref{fig:g1-vs-b}.}
\label{fig:gsc-vs-b}
\end{figure}

A necessary condition for monopole condensation is
$G_1 \leq \ln 7$. This is achieved for $b \gtrsim 0.27$ fm in full QCD and
for $b \gtrsim 0.35$ fm in the quenched theory. In
Fig.~\ref{fig:g2-g3-vs-b} we show the coupling constants $G_2$ and $G_3$.
We see that in full QCD $G_2$ is systematically smaller than in the quenched
theory for all values of $b$, as in the case of $G_1$,
while $G_3$ shows the opposite behavior. The other coupling constants show
little difference between full and quenched QCD. In 
Fig.~\ref{fig:g28-g29-vs-b} we show, as an example, $G_{28}$ and $G_{29}$.
Because the magnetic charge of the monopoles is in almost all cases $\pm 1$, 
and $G_1$ is the dominant coupling constant,
the monopole action can be approximated by $G_1\, L$, where $L$ is the
length of the monopole loop. As a result, a smaller self-coupling $G_1$ will 
give rise to a larger value of $L$ and a larger monopole density, and vice 
versa. This is to say that both observations, namely that the monopole density
increases and the self-coupling decreases with decreasing quark mass, are 
consistent with each other.

To shed further light on the dynamics of the monopoles we have looked at the
coupling $G$ of two ($n=1$) parallel monopole currents,
$k(s,\mu)$ and $k(s^\prime,\mu)$,
as a function of the distance $d=\sqrt{\sum_\mu(s_\mu-s^\prime_\mu)^2}$
between them.
In Fig.~\ref{fig:comp} we show $G$ together with the lattice Coulomb 
propagator. We
see that at distances $d \gtrsim 2$ the interaction becomes weaker than
Coulomb in both full and quenched QCD. This is consistent with the screening
effect discussed in Section 3. We do not see any difference between full QCD 
and quenched QCD though.

We may parameterize the effective monopole action by
\begin{equation}
S(k)= S_{\rm Coulomb}+S_{\rm self}+S_{\rm 4-point}+S_{\rm 6-point}
+S_{\rm add},
\end{equation}
where
\begin{equation*}
S_{\rm Coulomb} = \frac{g_{\rm m}^2}{2}
\sum_{s,s'} \sum_{\mu =1}^4 k(s,\mu)\Delta^{-1}(s-s') k(s',\mu)\,,
\end{equation*}
\begin{equation}
\begin{split}
S_{\rm self} &=  G_{\rm self} \,S_1(k) \,,\\[0.5em]
S_{\rm 4-point} &= G_{\rm 4-point} \, S_{28}(k)\,,\\[0.5em]
S_{\rm 6-point} &= G_{\rm 6-point} \, S_{29}(k)\,,
\end{split}
\end{equation}
\noindent
and $S_{\rm add}$ includes 12 additional two-point interaction terms.
In Fig.~\ref{fig:gsc-vs-b} 
we show the self-coupling $G_{\rm self}$ and the Coulomb coupling $g_{\rm m}$.
It is interesting to see that in full QCD $G_{\rm self}$ is smaller than in the
quenched case, while the Coulomb coupling is almost unchanged.
Corrections to the Coulomb interaction are found to be very small in the 
infrared region.

\section{Conclusions}

We have performed a detailed study of the dynamics of the QCD vacuum, thereby
focussing on the abelian degrees of freedom in the MAG. Our main objective was 
to find out how the vacuum reacts to the introduction of dynamical color 
electric charges (quarks). 
The monopole density was found to increase by more than a factor of two if
we decrease the quark mass from $m_\pi/m_\rho = 1$ (the quenched limit) to 
$m_\pi/m_\rho \approx 0.6$, both for the total
number of monopoles and for the monopoles in the infrared clusters. Related 
to that, we found that the magnetic screening length decreased by 30\% over
that range. 
The string tension, the static potential and the structure of the flux tube,
on the other hand, remained almost the same.
We verified the dual Amp\`ere law in full QCD and in the pure $SU(3)$ gauge 
theory. This result lends further support to the dual superconductor model of 
the vacuum in full and quenched QCD.
The width of the abelian flux tube was found to be $\delta=0.29(1)$ fm in 
both cases.
Another characteristic feature of the flux tube is the penetration length. 
We obtained $\lambda = 0.15(1)$ fm in full QCD and $\lambda = 0.17(1)$ fm 
in the quenched case. This results in a dual photon mass of $1.3(1)$ GeV and 
$1.2(1)$ GeV, respectively.
Decomposing the abelian gauge field into monopole and photon parts allowed
us to study flux tubes up to a length of $\approx 1.6$ fm in full QCD. No
signal of string breaking was found. Comparing flux tubes of various lengths
$R$, it turned out that the width of the flux tube does not depend on $R$,
contrary to the prediction of the Nambu-Goto effective string theory.
The effective monopole action was determined. In full QCD  
the monopole self-coupling was found to be systematically smaller than in the 
quenched theory. The main contributions to the effective monopole action are
found to be the self-interaction and the Coulomb interaction.

\section*{Acknowledgements}

We wish to thank the UKQCD collaboration for letting us use their 
dynamical gauge field configurations at $\beta=5.2$, and A. Irving for 
computing $r_0/a$ on all configurations. Furthermore,
we would like to thank G.~Bali, A.~Hart, T.~Kovacs, M. M\"uller-Preussker and
H.~Suganuma for useful discussions. H.I. thanks the Humboldt University and 
the Kanazawa University for hospitality. V.G.B. acknowledges
support from JSPS RC 30126103. The numerical calculations have been done on 
the COMPAQ Alpha Server ES40 at Humboldt University 
and on the SX5 at RCNP, Osaka University. 
T.S. is supported by JSPS Grant-in-Aid for Scientific Research on Priority 
Areas 13135210 and (B) 15340073.
M.I.P is partially supported by grants RFBR 02-02-17308, 
RFBR 01-02-17456, INTAS-00-00111, DFG-RFBR 436 RUS 113/739/0, 
RFBR-DFG 03-02-04016, and CRDF award RPI-2364-MO-02.

\end{document}